\documentclass[]{aa}

\usepackage{graphicx}
                                     
\usepackage{natbib}  
\bibpunct{(}{)}{;}{a}{}{,}

\usepackage{txfonts}  

\begin{document}

   \title{Star formation and bar instability  in cosmological haloes}

    \author{Anna Curir,\inst{1}  Paola Mazzei\inst{2} and Giuseppe Murante\inst{1}}

	\offprints{ A. Curir}

    \institute{INAF-Osservatorio Astronomico di Torino. Strada Osservatorio 20
 -10025 Pino Torinese (Torino). Italy.   e-mail:curir@to.astro.it
    \and
     INAF-Osservatorio Astronomico di Padova. Vicolo Osservatorio 5 - 35122
   Padova. Italy}
     \date{}

\abstract
{This is the third of a series of papers presenting the first attempt to
analyze the growth of the bar instability in a consistent cosmological
scenario. In the previous two articles we explored the role of the cosmology
on stellar disks, and the impact of the gaseous component on a disk embedded
in a cosmological dark matter halo.}
{The aim of this paper is to point out the impact of the star
formation on the bar instability inside disks having different gas
fractions.}
{We perform  cosmological simulations of the same  disk-to-halo mass systems
as in the previous works where the star formation was not triggered. We
compare the results of the new simulations with the previous ones to
investigate the effect of the star formation by analysing the morphology of the
stellar  components,
the  bar strength, the behaviour of the pattern speed. We follow the gas and
the central mass concentration during the evolution and their impact on the
bar strength.}
{In all our cosmological simulations a stellar bar, lasting 10 Gyr,
is still living at z=0.\\
The central mass concentration of gas and of the new stars has a mild action
on the ellipticity of the bar but is not able to destroy it; at z=0 the
stellar bar strength is enhanced by the star formation.
The bar pattern speed is decreasing with the disk evolution.}
{}
        \authorrunning{Curir, Mazzei \& Murante}
     \titlerunning{Star formation in barred  disks}

 \maketitle

\section{Introduction}

The connection between the bar feature and the star formation  process has
already been pointed out in the past: \citet{mart77} observed that
non interacting galaxies  displaying the highest star forming activity have strong and long bars.
Conversely not all strong and long bars are intensively creating stars.
\citet{Ma01}  performed SPH simulations to investigate the dependence
of the star formation in disks on the geometry and dynamical state of the DM
halo.
They showed that the star formation lengthens the life time of the bar.
However in  their works the evolution of the disk+halo system arises in a isolated framework, 
outside the cosmological scenario.\\
The treatment of the star formation in galaxies involves
a large number of physical processes  which arise with different time
scales  and  spatial lengths.
Therefore its inclusions in simulations of galaxy formation is complex
(\citet{MaCu03} and references therein) and many sub-grid parametrisations
(i.e. representing all the astrophysical processes that cannot be resolved due
to resolution or computing power limitations) of the star formation process, exist
in literature.
To mimic all these processes with a numerical model of a realistic disk
galaxy is difficult too. Thus the numerical work on this subject focuses
either on a detailed analysis of the interstellar medium (ISM) and a smaller simulated area
\citep{Wa01, avil2000} or on the study of the global instabilities and of the star formation
at the cost of simplified ISM models \citep{Spri04,  Sem02, Comb02}.
Models that have tackled  both  topics have either been in two dimensions
\citep{WaNo01} or restricted to a box size a few hundred parsecs across
\citep{Wa01}. \citet{Ros95} worked in two dimensions allowing the ISM to evolve
self-consistently but treated the stars as collisional rather than collisionless fluid.
\citet{task06} presented the first three dimensional simulations of a global disk without
 the need to simplify the structure of the ISM. Their paper is devoted to understand the
fundamental processes  of star formation and feedback in a disk galaxy.\\
Here we  performed cosmological simulations to analyse the effects of the
star formation process inside stellar-gaseous disks embedded in a DM cosmological halo evolving
in a fully consistent scenario.
We employ the sub-grid star formation model by \citet{Spri03} which is able to
produce a self-regulated star formation in galactic disks.
We point out that our model is not a general galaxy evolution model, since the
gradual formation and growth of the stellar disk has not been taken into
account. However our approach,  developed in two previous papers \citep{Cu06, Cu07},
allows us  to investigate the effects of different parameters like the
disk-to-halo mass ratio and the gas fraction inside the disk, on the growth
of the bar instability, its coupling with the star formation rate  and  its dependence
on such parameters in a self-consistent cosmological framework.
We compare the results of  such a new set of simulations with those of  our previous sets
with the star formation switched off (\citet{Cu06}, hereafter
Paper 1, \citet{Cu07}, hereafter Paper 2). In Paper 1 we presented
simulations  of purely stellar disks with the same disk-to-halo mass ratios,
in the same cosmological scenario, and with the same initial conditions as in
this new paper. In Paper 2 we investigated the growth and the evolution of the bar
instability in disks with the same disk-to-halo mass ratios as in Paper 1, and
different gas fractions, without star formation.
Thus, we will compare the results of simulations here performed, with those
corresponding to the same disk-to-halo mass ratio as in Paper 1 and to the same
gas fraction as in Paper 2.
The comparison between the parameters characterising the bar can be done only
evaluating that of the old stars component, since the new stars are not
present in Paper 1  and  2. 
In both such Papers it was shown that in DM dominated disks a bar
feature is triggered and maintained by the Cosmology whereas, in the more
massive disks, a gas fraction 0.2 is able to destroy the bar.
The focus of the present
work is to determine if the  star formation changes such a result.

The plan of the paper is the following: in Section 2, we  summarise
our recipe for the initial $disk+halo$ system fully described in Paper 1,  and
present our star formation recipe.
In Section 3  there are our cosmological simulations, in Section 4 we point
out our  results.
The parameters related to the bars formed in the new and the old stars and to the
global, old+new, stellar populations are given in such a
section.
Section 5 is devoted to our discussion and conclusions. It contains
a Table including the parameters characterising the old stellar bar at z$=0$
in the  three Papers to allow a more easy comparison between the results of  our
Papers (Table 2).

\section{ Method}

We embed a gaseous and stellar disk inside a  cosmological halo selected in a
suitable slice of Universe and follow its  evolution inside a cosmological
framework: a $\Lambda$CDM  model with
$\Omega_{m}=$0.3, $\Omega_{\Lambda}=$0.7, $\sigma_8=$0.9, 
$h=$0.7, where $\Omega_{m}$ is
the total matter of the Universe, $\Omega_{\Lambda}$ the
cosmological constant, $\sigma_8$  the normalisation of the power spectrum,  
and $h$ the value of the Hubble constant in units of 100$h^{-1}$
km\,s$^{-1}$Mpc$^{-1}$. 

A detailed description of our method to produce the cosmological
scenario where the disk is evolved has been  given in Paper 1.
Here we present a short  overview of our  recipe.
Paper 1 showed that the numerical resolution does not
impair  their main result: in pure stellar disks, long living bars are a 
'natural' outcome of the cosmological scenario.  

\subsection{The DM halo}

To select the DM halo, we perform a low-resolution simulation of a {\it
concordance} $\Lambda$ CDM cosmological model, starting from redshift 20.\\
With a standard 'friends of friends algorithm' we selected one suitable DM
halo with a mass M\( \sim \)10\( ^{11} \)\( h^{-1} \) M\( _{\odot } \) (at
z=0). We resample it with the multi-mass technique described in
\citet{Kly01}. The particles of the DM halo, and those belonging to a sphere
with a radius $4 h^{-1}$ Mpc, are followed to their Lagrangian position and
re-sampled to an equivalent resolution of 1024 \(^{3} \) particles.  The total
number of DM particles in the high resolution corresponds to a DM mass
resolution of $ 1.21\, 10^{6} h^{-1}\,M_{\odot }$.  The high-resolution DM
halo is followed to the redshift z=0.  We run the DM simulation, to extract
the halo properties in absence of any embedded stellar disk.  The mass of our
halo at z=0, \(1.03\cdot 10^{11}h^{-1} \) M\( _{\odot } \), corresponds to a
radius, \( R_{vir} = 94.7h^{-1}\) Kpc, which entails 84720 halo particles.
The selected halo is living in an under-dense environment. From its accretion
history (see Fig. 1 in Paper 1) we conclude that our halo undergoes no
significant merger during the time it hosts our disk, nor immediately before.
The halo density profile is well--fitted by a Navarro, Frenk and White (NFW)
form (\citet{Nav96}; \citet{Nav97}) at z$ \leq$ 2. The concentration,
C$_{vir}$, equal to R$_{vir}/R_s$, takes an high value, 18.1, confirming that
this halo does ``form'' at quite high redshift ( e.g. \citet{Wec02} for a
discussion about the link between concentration and assembly history of the
halo).  The dimensionless spin parameter of the halo is 0.04 at z=2, near to
the average one for our cosmological model \citep{Mal02, Bet07}.\\ 
Cosmological
DM haloes are not spherical  \citep[ and references therein]{Bet07}.
From a recent analysis of the halo shapes from
the Millennium simulation, \citet{Bet07} show that halos
 are triaxial and prolate. Their shape is driven by the hierarchical 
formation. 
The prolateness of our halo at z=2, where R$_{vir}=$30\,Kpc, is 0.9, quite the same as at the disk radius (Paper 1,
Table 1).

\subsection{The baryonic disk}

The spatial distribution of  particles  follows the exponential surface density
law: $\rho =\rho_0\exp -(r/r_0)$ where r$_0$ is the disk scale length,
$r_0=4h^{-1}$\, Kpc, and $\rho_{0}$ is the surface central density. 
The disk is truncated at five scale lengths  with a radius:
R$_{disk}=$20$h^{-1}$\,Kpc. To obtain each disk particle's position according
to the assumed density distribution, we used the rejection method \citep{Pre86}.
We used 56000 star particles and 56000 gaseous particles to describe our disk.
The (Plummer equivalent) softening length,  the same for DM, gas, and star particles,
is $ 0.5 h^{-1}$\,Kpc in comoving coordinates.

 We embed the disk in the high resolution cosmological  simulation, at 
 redshift 2, in a plane perpendicular to the  angular momentum vector of
 the halo and in gravitational equilibrium with the potential.  Its centre of 
mass corresponds to the  minimum potential well of the DM halo.
The  initial redshift corresponds to 10.24 Gyr down to $z=0$ in our chosen
cosmology.
During the evolution, new star particles are formed from the gaseous
particles. 
We will refer to such new component as 'new stars component', whereas we will
call 'old stars component' the non dissipative particles present in the disks
at z=2.

\subsection{Star formation recipe}

We use the sub-grid star formation prescription by Springel \& Hernquist
(2003). In such prescription, when a gas particle overcomes a given density
threshold its gas content is considered to reside in a multi-phase state,
corresponding to the  equilibrium solution of an analytical model
describing the physics of the multi-phase interstellar medium. Such a
solution gives an effective temperature for the gas particle, obtained as a
weighted average of the (fixed) temperature of the cold phase and that of the
hot phase. The temperature of the hot phase is set by supernova feedback
and by the efficiency of cloud evaporation. The resulting  inter stellar
medium (ISM) has an effective
equation of state $P_{eff}(\rho)$ which is stiffer than isothermal and 
prevent Toomre instabilities even when a large amount of gas is present.  When
this prescription is used, the star formation rate agrees with the Schmidt law
 in disk galaxies, as obtained e.g. by Kennicut (1998). A detailed description
of the numerical prescription and  several tests can be found in Springel
\& Hernquist (2003).
The effective temperature drives hydrodynamical interactions of the gas particles. Moreover, the model consistently gives a star formation rate, which
is used to spawn a star particle from the gas one on the basis of a stochastic
prescription; the initial mass function adopted is the Salpeter's one. The
star forming from the gas has the same position and velocity of the gaseous
particle.

\section{Simulations}

We performed six cosmological simulations of a disk+halo system.
We exploited the  parallel Tree+SPH N-body code GADGET-2 \citep{Spri05}
(courtesy of V. Springel).
The simulations run on the  CLX computers located at
the CINECA computing center (BO, Italy ) and on OATo Beowulf-class cluster
of 32 Linux-based PC  at the Osservatorio Astronomico di Torino.
\begin{table*}
\caption{ Simulations: final values.  o.s.=old stars , n.s.= new stars}
\label{cosmsimtable}
\centering
\begin{tabular}{c c c c c c c c c c c c }
\hline\hline
N  & M$_{disk}$ & gas fraction & $\epsilon$ (o.s.)&  $\epsilon$ (n.s.)  & Q$_b$ (o.s.) & Q$_b$ (o.s.+n.s.)  & a$_{max}$ (o.s.)
& a$_{max}$ (n.s.)&  bulge(o.s.)& bulge(n.s.)  &
bars in bars \\  
\hline
 c1 &  0.33 & 0.1 & 0.65 & 0.72 & 0.45 & 0.5 & 8 & 4 & y & n & n \\
 c2 &  0.33 &  0.2 & 0.55 & 0.55 & 0.51 & 0.66& 11 & 5.7& y& y & n \\
 c3 &  0.33 & 0.4  & 0.6 & 0.55& 0.58 & 0.69&  8.4 &6& y & y & n \\
 c4 &  0.1 & 0.1 & 0.39 &0.01 & 0.5  &0.41 & 3 & - & n &y & y \\
 c5 &  0.1 & 0.2 & 0.45 &0.03& 0.46  &0.42& 3  & - & n & y & y\\
 c6 &  0.1 & 0.6 &  0.5 &0.02&  0.51 & 0.47 &3& - & n & y& y \\
\hline
\end{tabular}
\end{table*}
The main parameters and the final properties of our set of  simulations
are listed in Table  \ref{cosmsimtable}.
We define as a geometrical measure of the bar strength the value of the
ellipticity, $\epsilon=1-b/a$ (Table \ref{cosmsimtable}); a strong bar
corresponds to $\epsilon \geq 0.4$.  A more dynamical measure of the bar
strength at radius R has been defined by \citet{Comb81} by using the
parameter: $Q_t= {F_T^{max}(R)\over{<F_R(R)>}}$ where $F_T^{max}(R)=[{\partial
\Phi(R,\theta)}/{\partial \theta}]_{max}$ is the maximum amplitude of
tangential force and $<F_R(R)>=R({\partial \Phi_0}/{\partial r})$ is the mean
axisymmetric radial force derived from the $m=0$ component of the
gravitational potential at the same radius, R. The maximum value of $Q_t(R)$ provides a measure of the
bar strength $Q_b$ for the whole galaxy. Stronger bars correspond to higher Q$_b$ values.  In order to compare the results with
the Papers 1 and 2 the values of $Q_b$ are evaluated in the bars formed in the
old stars component, since in our previous works we didn't  include
the star formation process.\\  We  give here in Table \ref{cosmsimtable} also the
value of the parameters related to the bars formed in the new stars component and to the bar
strenght of the global old+new stars populations.
In Table \ref{cosmsimtable} we present: the simulation
number (I column), the mass of the disk in code units (i. e. $5.9 \times
10^{10}\, M\odot$) (II col.), the fraction of gas, i.e. gas--to--disk mass
ratio (III col.)  Moreover we present, as final values (i. e.  at $z=0$) the
maximum ellipticity for the old stars system(IV col.), and for the new stars
(V col.) the bar strength
according to \citet{Comb81} for the old stars (VI col.), and for the old+new
stars system (VII col.), the major axis (physical Kpc)
corresponding to the maximum ellipticity for the old stars (VIII col.)and for
the new stars (IX col.), the morphology of the
inner region of the disk for the old stars (X col.) and for the new ones (XI col.), peculiar features inside the disk (XII
col.)

\section{Results}

In all our cosmological simulations, 
a stellar bar is still living at z=0 in the old star component.
The new stars component at z=0 is arranged in a bulge component which can or
cannot present a barred shape depending on the initial mass of the disk and on
the gaseous fraction.\\
In Figs 1-8 we  present  the  
isodensity contours of the  old stars,  gas and new stars for simulations in Table
\ref{cosmsimtable}.  All such figures have been built up with the same
box-size, number of levels, and density contrast (see caption of Fig.
\ref{evol1}),  as in Papers 1 and 2.\\

\subsection{ Morphologies}\label{morf}

\begin{figure}
\centering
\includegraphics[width=7cm]{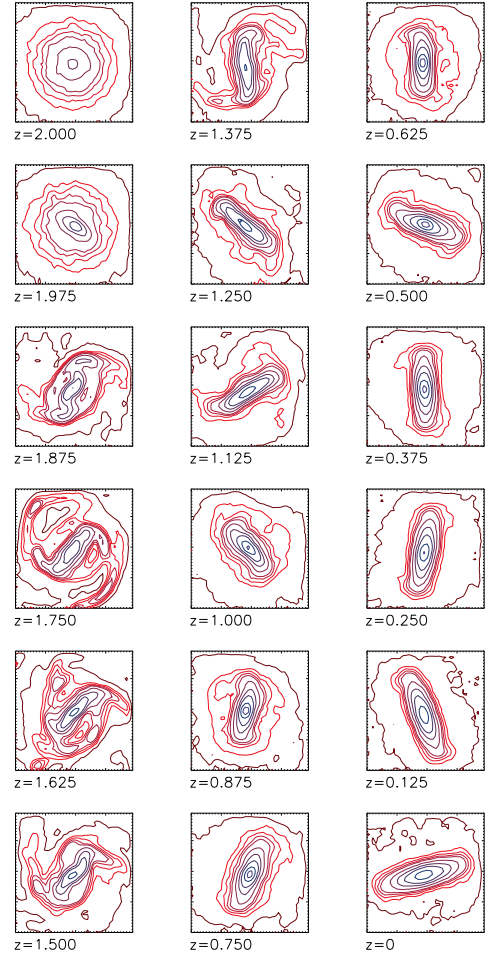}
\caption{ Face-on isodensity contours 
of simulation c1 from z=2 to z=0. Spatial resolution  is always 
0.5$h^{-1}$ {\it physical \,} Kpc and the
box size is 40 times the spatial resolution.
Contours are computed at 11 fixed levels ranging from  
$2\times 10^{-4}$  to $0.015$  in term of density  fraction of stars  
(gas) within the  spatial resolution to the total star (gas) density in the map.}
\label{evol1}
\end{figure}

\begin{figure}
\centering
\includegraphics[width=7cm]{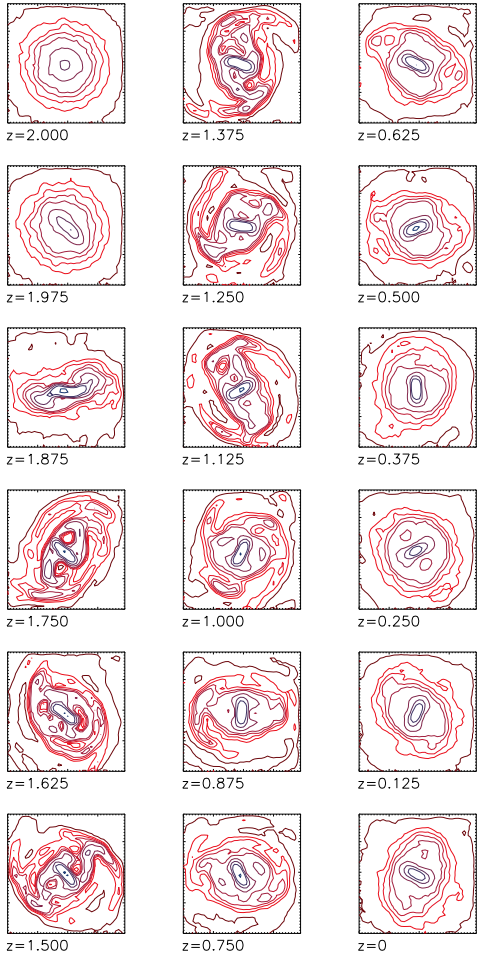}
\caption{ Face-on isodensity contours 
of simulation c4 from z=2 to z=0. Isodensity levels and spatial resolution are
the same as in Fig. \ref{evol1}}
\label{evol2}
\end{figure}

\begin{figure}
\centering
\includegraphics[width=7cm]{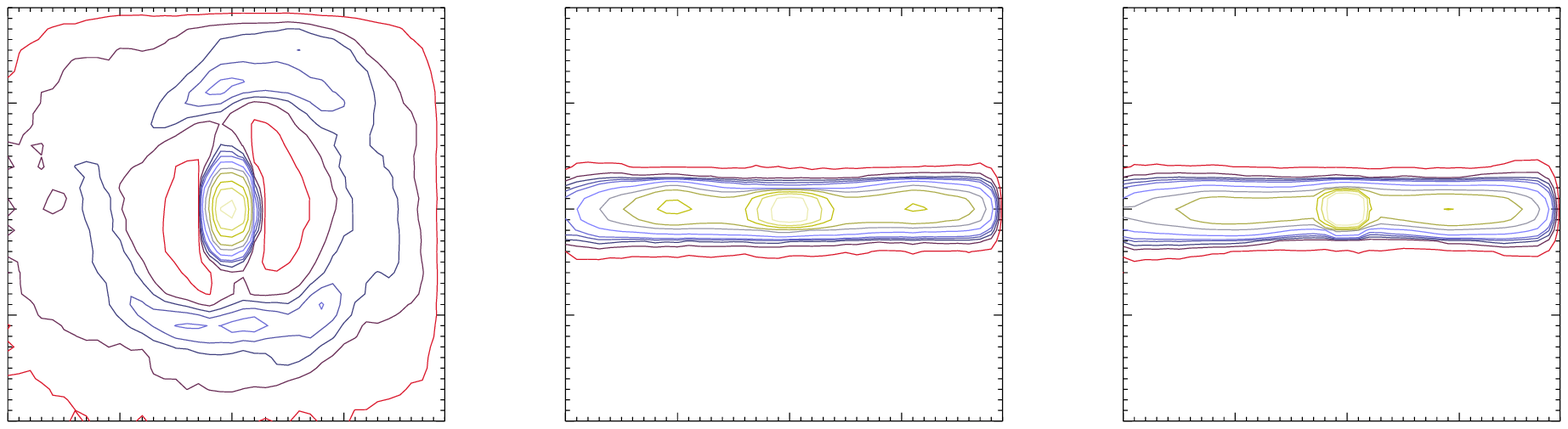}
\includegraphics[width=7cm]{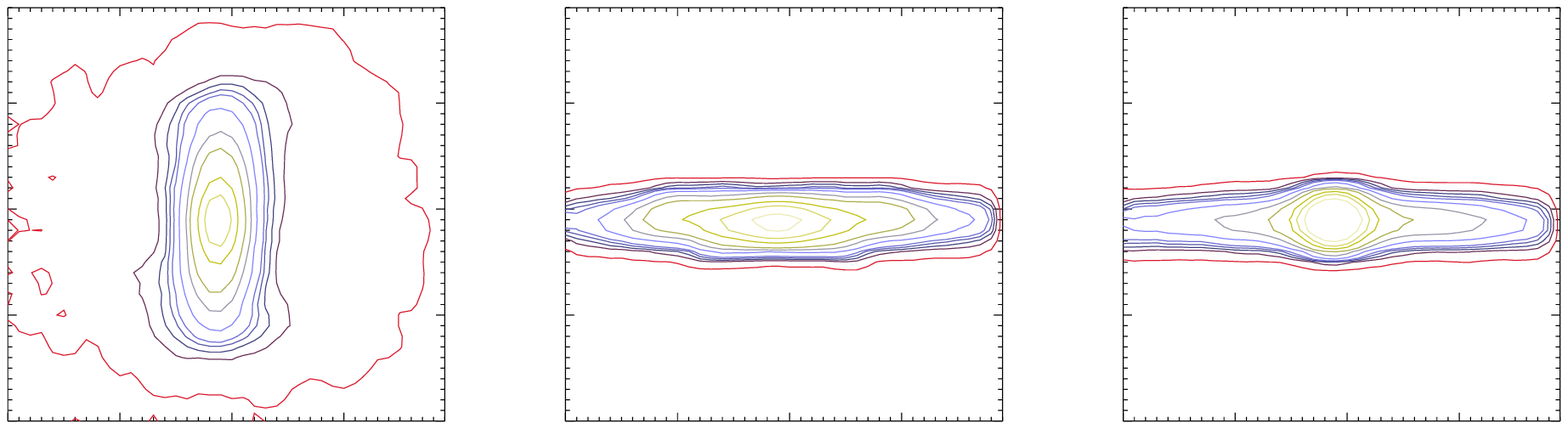}
\includegraphics[width=7cm]{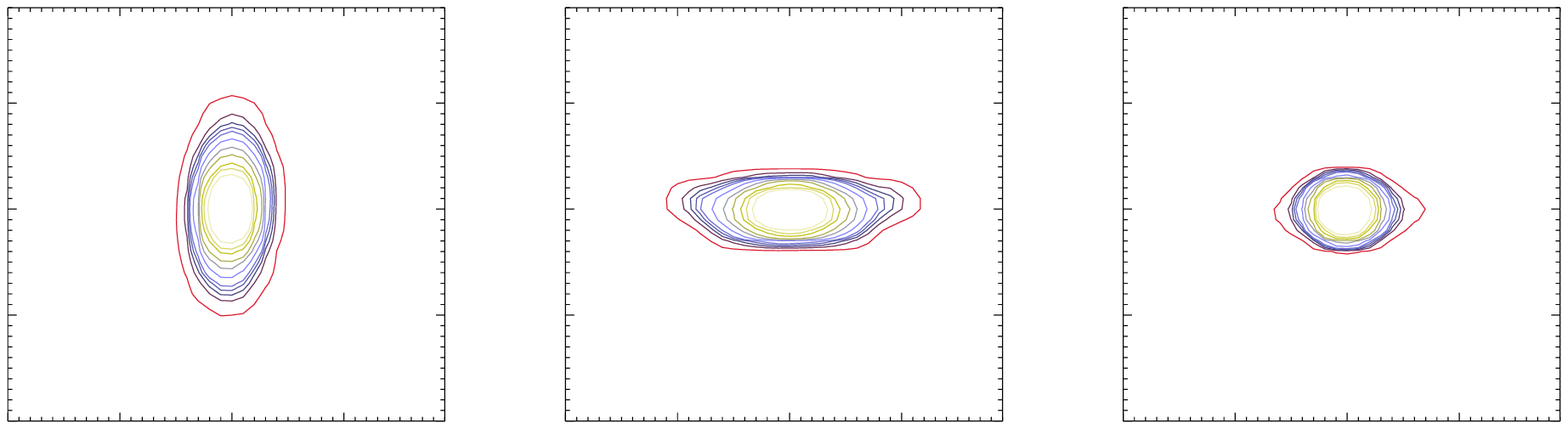}
\caption{Face-on, edge-on and side-on isodensity contours (from left to right)
of simulation c1 at z=0 (see text); top panel shows the gaseous component,
 middle panel the
old star component, bottom panel the new formed stars. Isodensity levels and
spatial resolution for gas and stars  are as in Fig. \ref{evol1}.}
\label{dens1}
\end{figure}

We will discuss first the bar features as far as the old star component is
concerned. 
 Fig.\ref{evol1} and Fig.\ref{evol2}  show the evolution of this
component from z=2 to z=0 for simulation c1 and c4. The bar feature is well defined
since z=1.75 in both the cases.  In the more massive case, the evolving bar 
becomes stronger and
dominates the disk, whereas in the less massive case, the bar remains a
central structure inside the disk, generating a bar-in-bar feature.
The strongest  bar arises from simulation c1 (Fig. \ref{dens1}). 
In this case,
the ellipticity, evaluated through  
the  isodensity plots, is 
higher (0.65) than the one measured    
in the purely  stellar disk  of the same mass (0.52, Paper 1), but slightly
smaller than that measured in the  case of a disk with the same mass and the same
gas fraction (0.68, Paper 2).
The enhancement of the stellar bar strength in comparison to the   gas
  free case is due
to the   coupling of the gaseous component to the stellar
one: the gaseous bar is indeed superimposed to the stellar one, as pointed out 
in Paper 2.  
However, this effect is slightly reduced by the star formation activity. 
\begin{figure}
\begin{center}
\includegraphics[width=7cm]{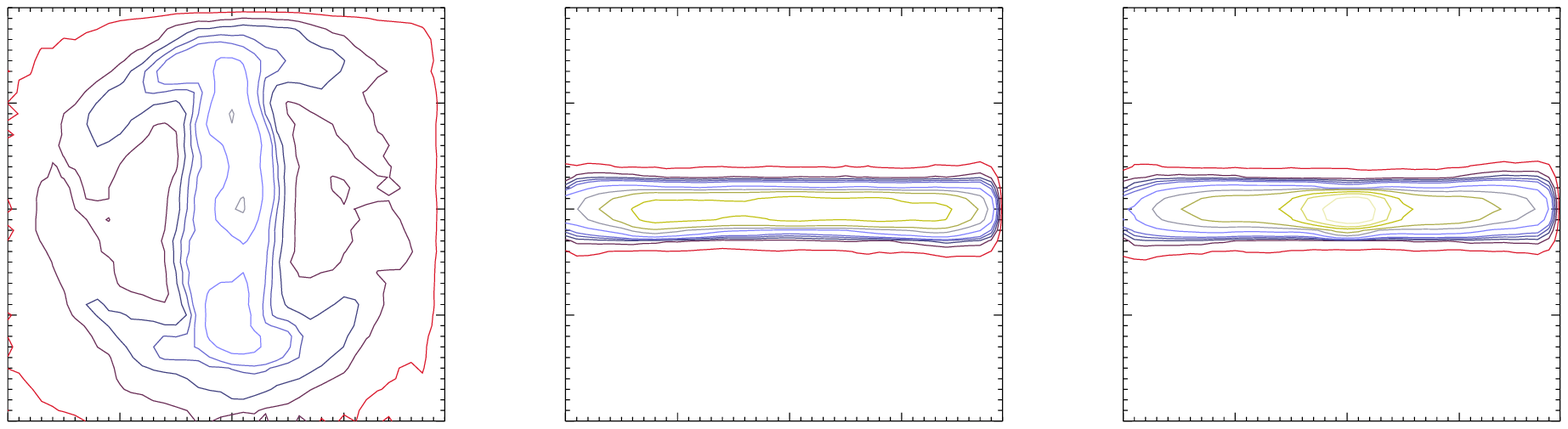}
\includegraphics[width=7cm]{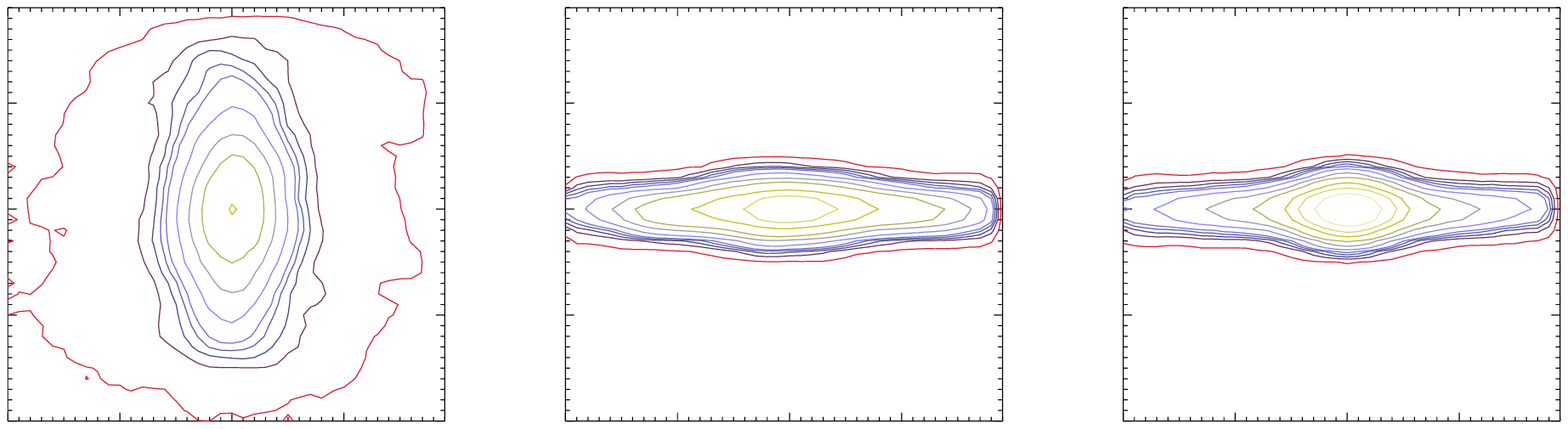}
\includegraphics[width=7cm]{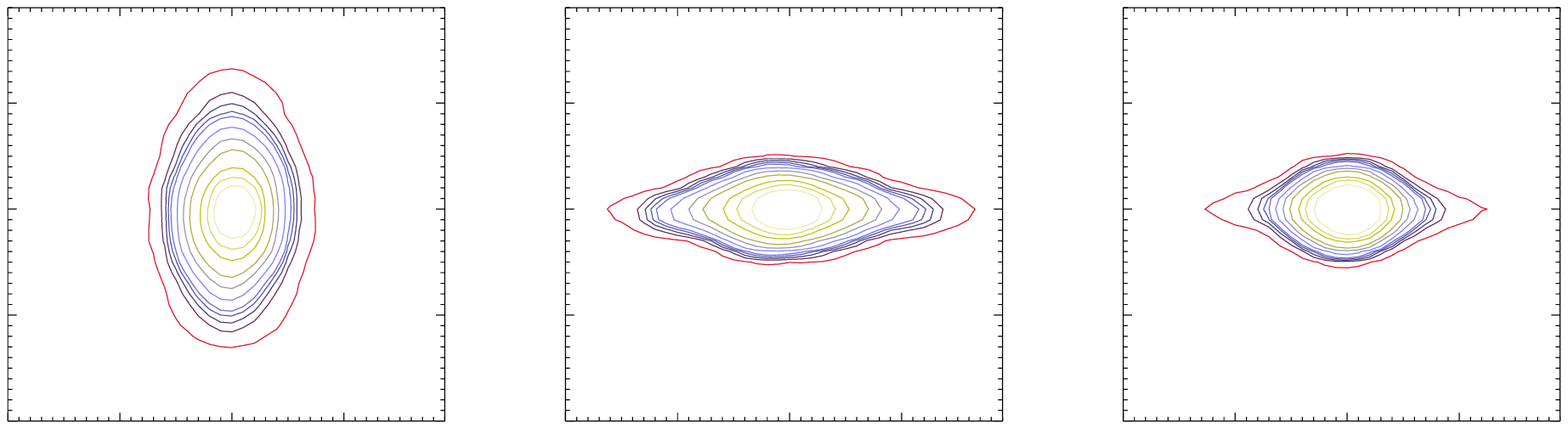}
\end{center}
\caption{Same as in Fig. \ref{dens1} but for simulation c2}
 \label{dens2}
\end{figure}
\begin{figure}
\begin{center}
\includegraphics[width=7cm]{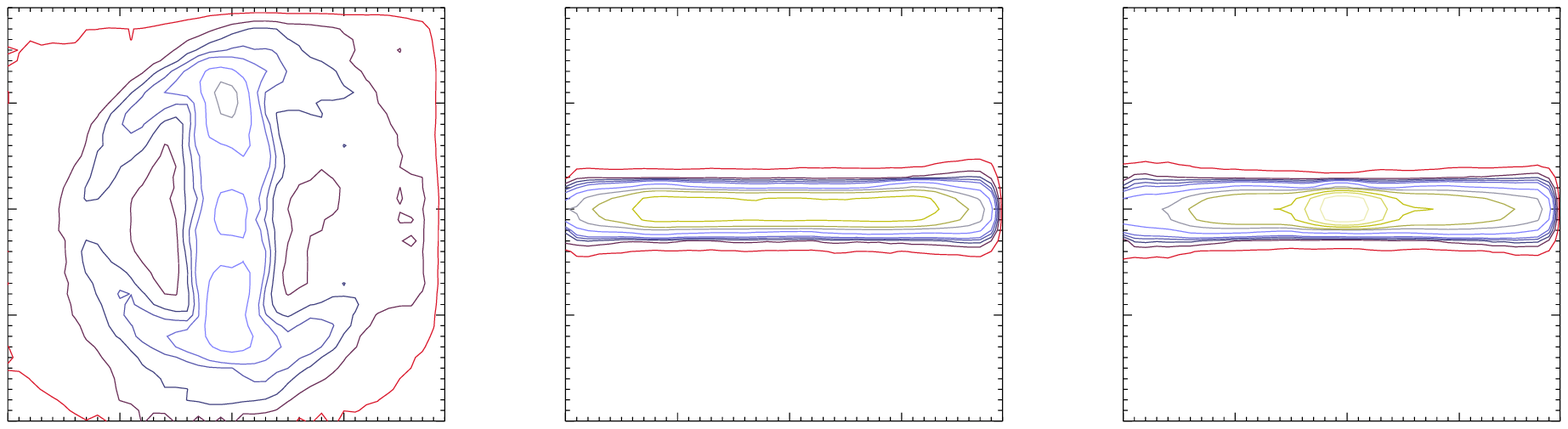}
\includegraphics[width=7cm]{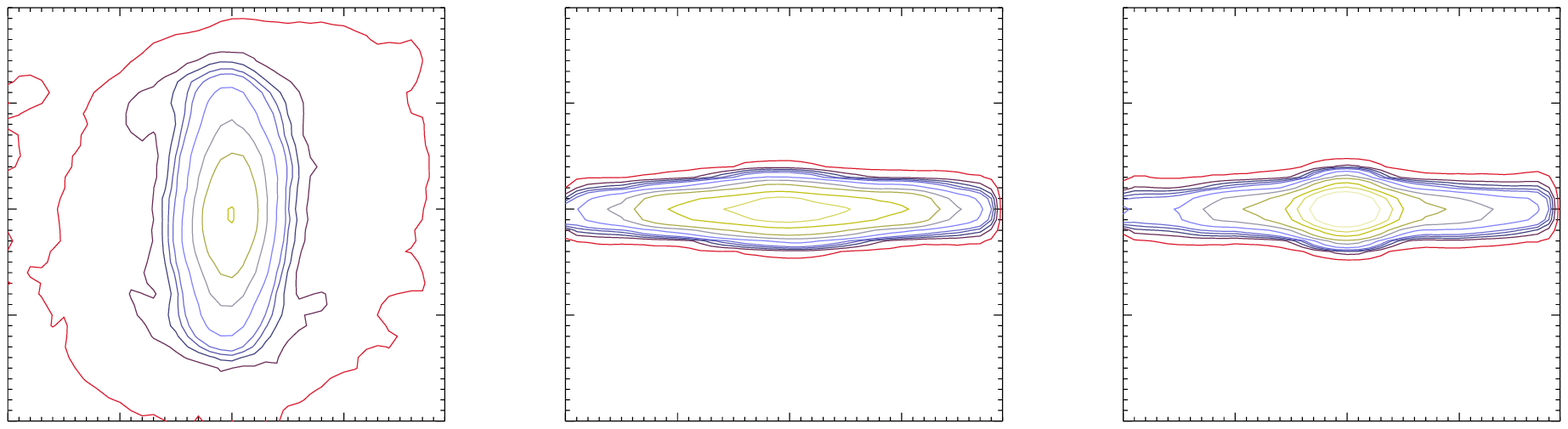}
\includegraphics[width=7cm]{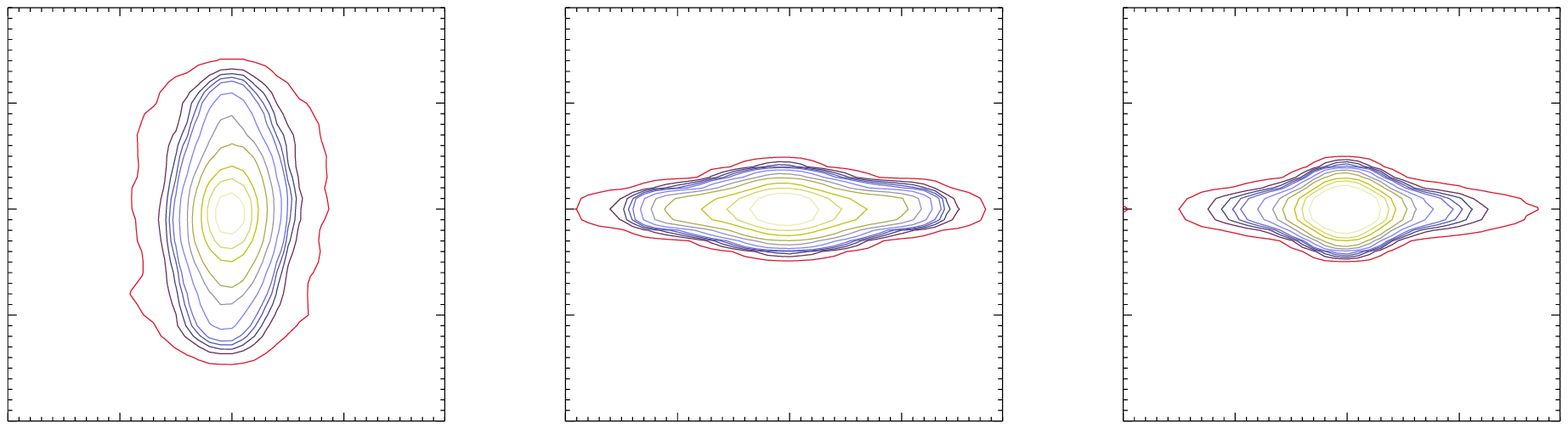}
\end{center}
\caption{Same as in Fig. \ref{dens1} but for simulation c3}
\label{dens3}
\end{figure}
\begin{figure}
\begin{center}
\includegraphics[width=7cm]{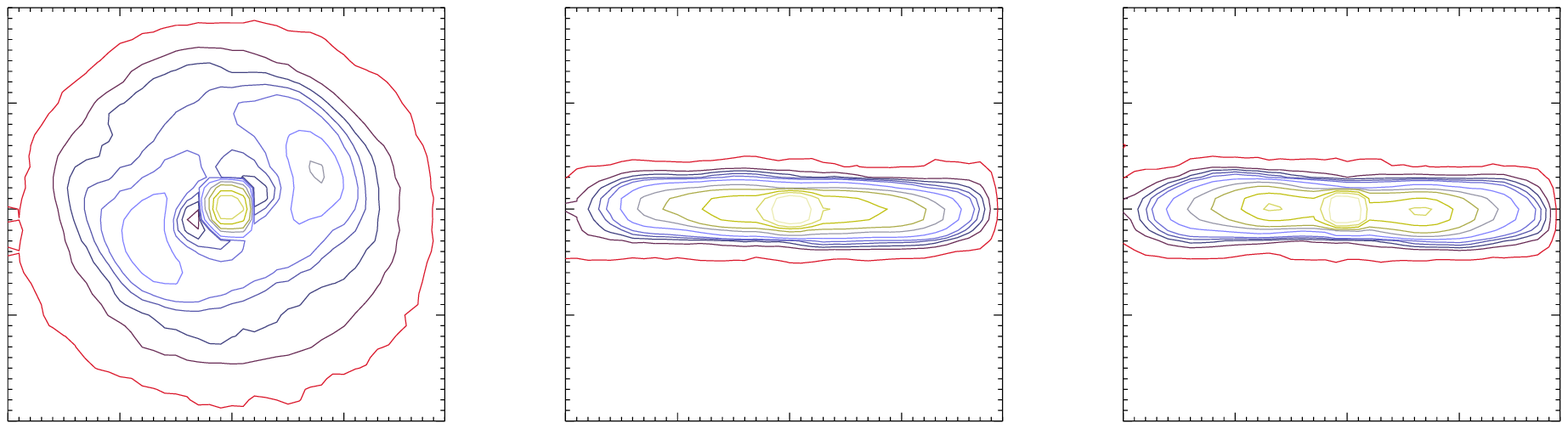}
\includegraphics[width=7cm]{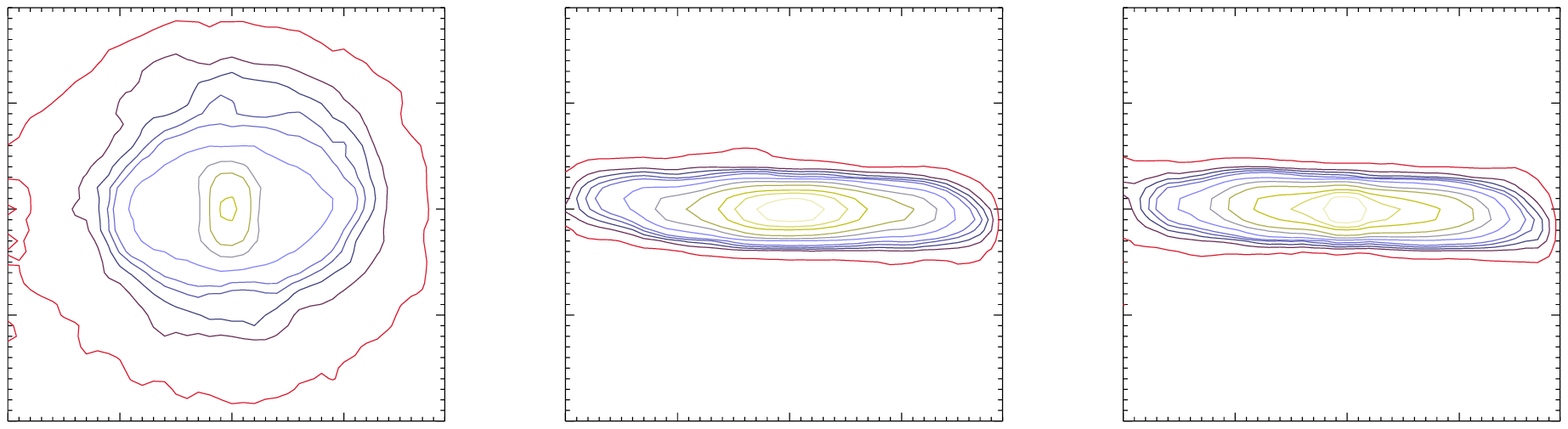}
\includegraphics[width=7cm]{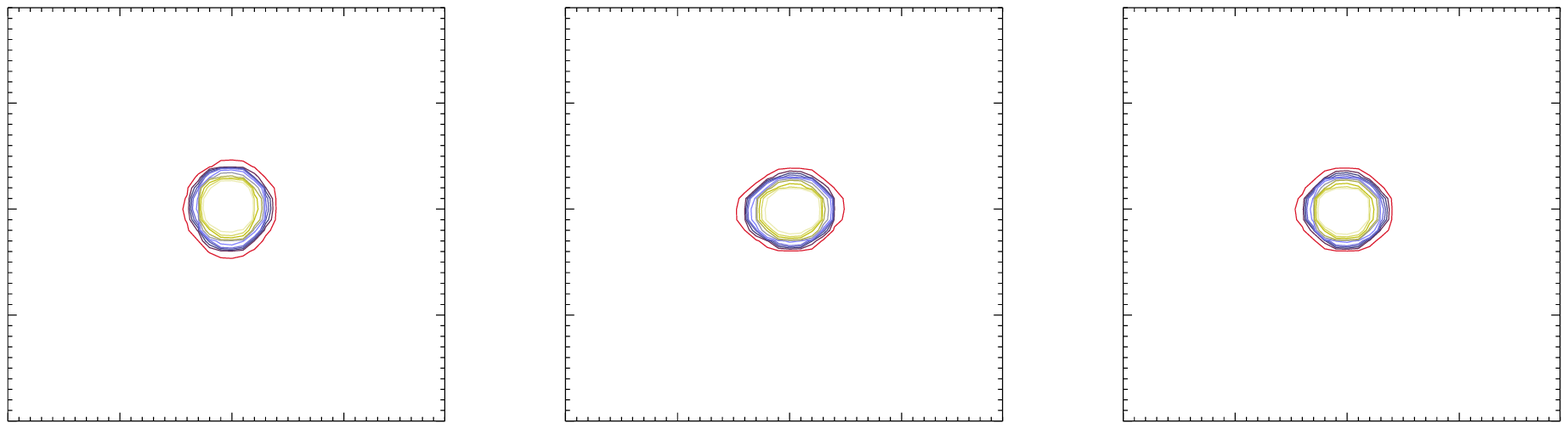}
\end{center}
\caption{Same as in Fig. \ref{dens1} but for simulation c4}
\label{dens4}
\end{figure}
\begin{figure}
\begin{center}
\includegraphics[width=7cm]{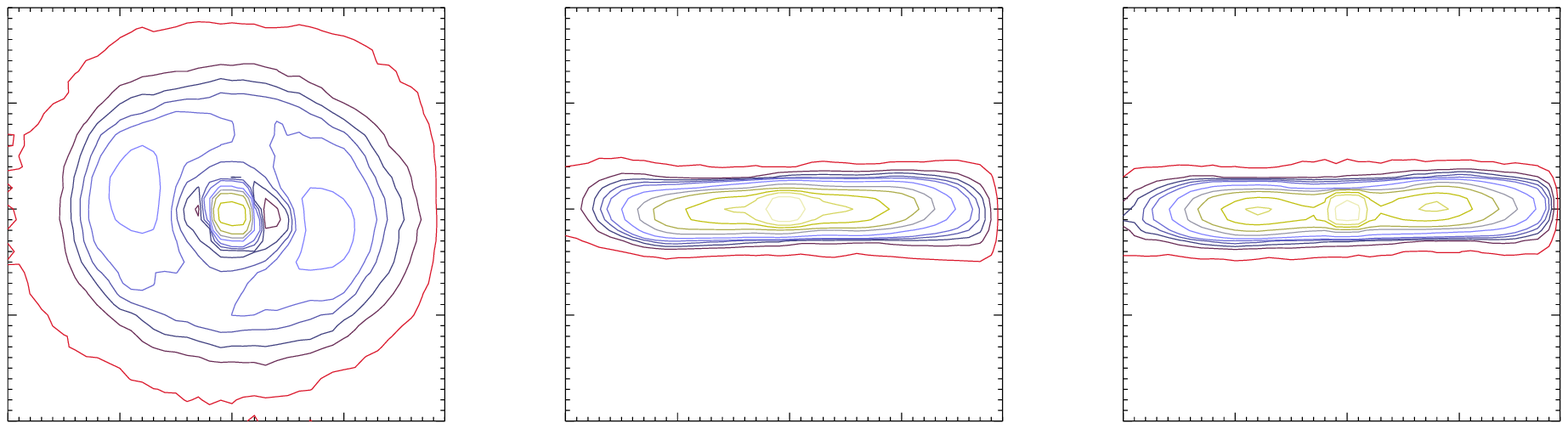}
\includegraphics[width=7cm]{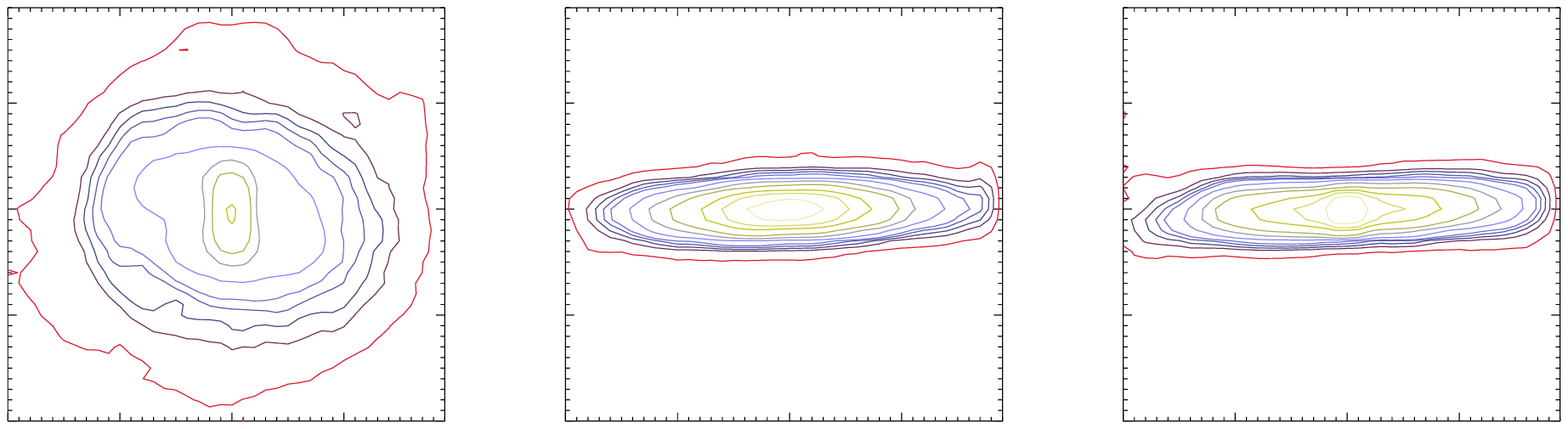}
\includegraphics[width=7cm]{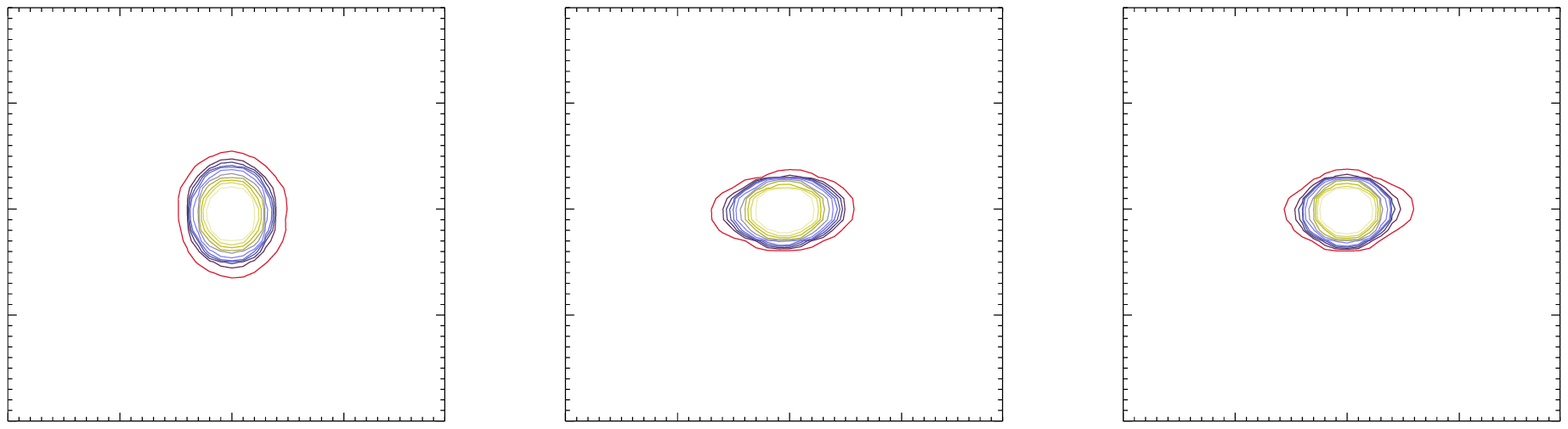}
\end{center}
\caption{As in Fig. \ref{dens1} for  simulation c5.}
\label{dens5}
\end{figure}
\begin{figure}
\begin{center}
\includegraphics[width=7cm]{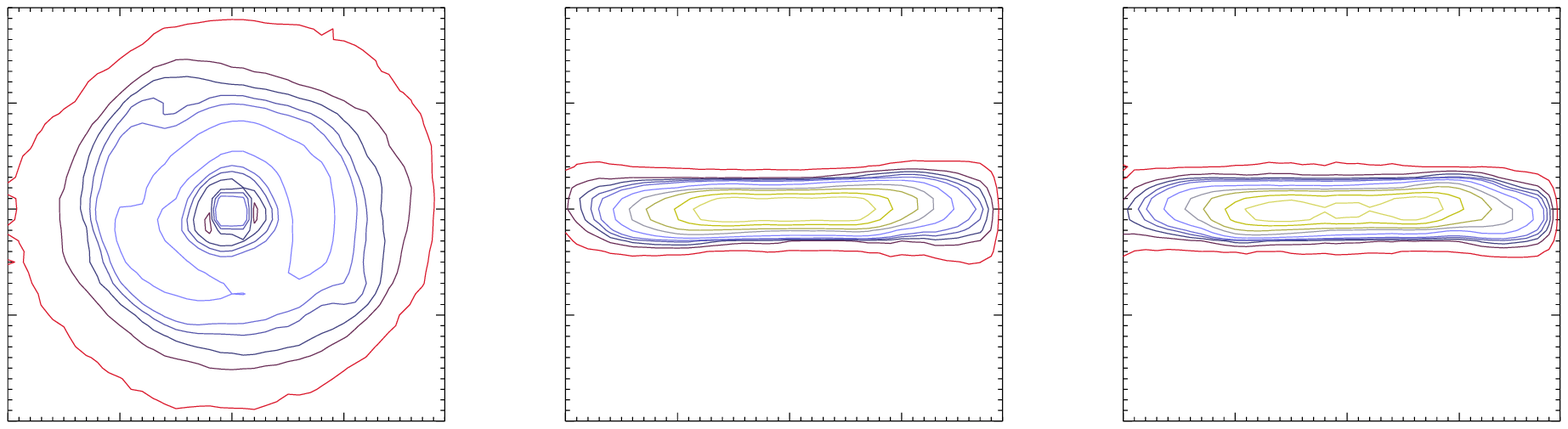}
\includegraphics[width=7cm]{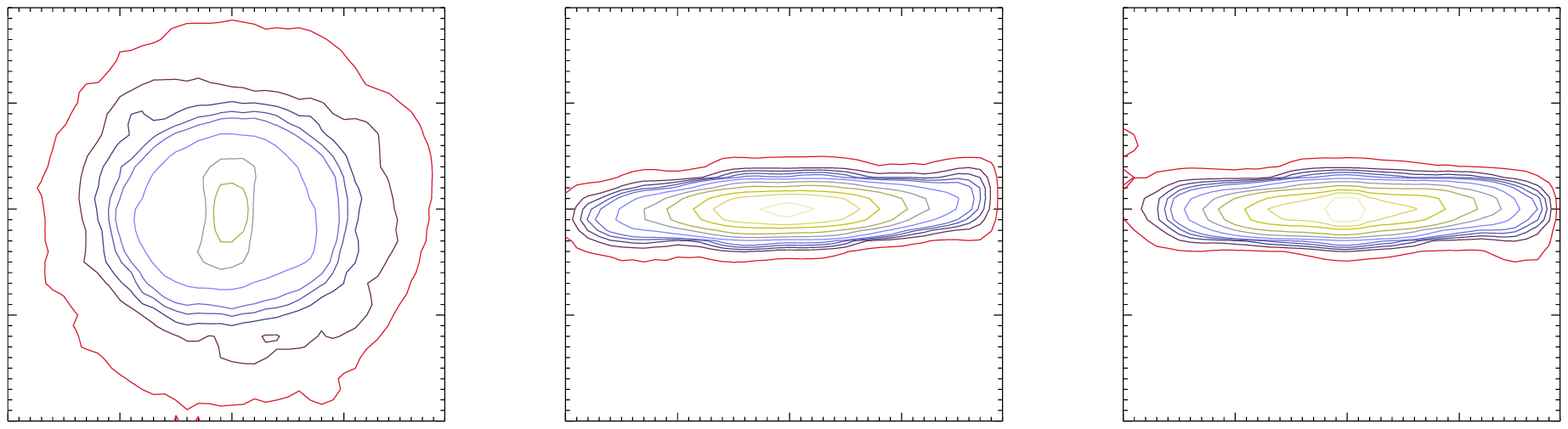}
\includegraphics[width=7cm]{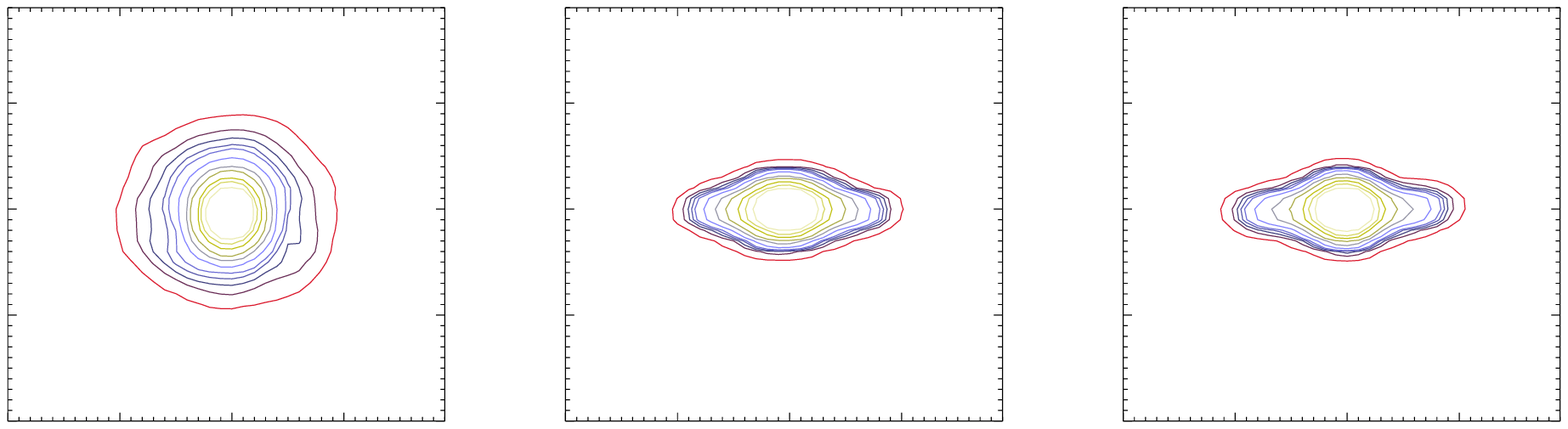}
\end{center}
\caption{As in Fig. \ref{dens1} for  simulation c6.}
\label{dens6}
\end{figure}
\begin{figure}
\begin{center}
\includegraphics[width=7cm]{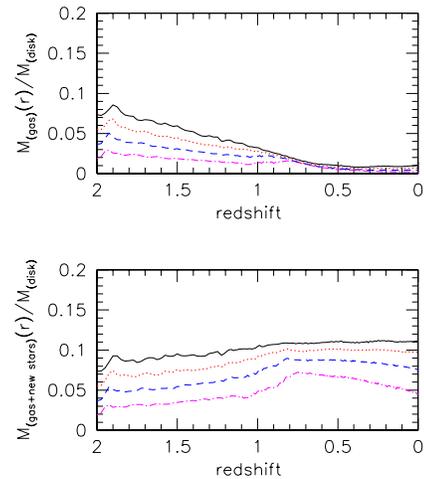}
\end{center}
\caption{Gas--to-- disk mass ratio at different disk radii :  2 Kpc (magenta
  dashed-dotted line), 3 Kpc
  (blue dashed  line), 4  Kpc (red dotted line), 5 Kpc (black full line) for 
simulation c2.
In the upper panel we present the time behaviour of  the gas as fraction of the
disk mass, in the lower panel we see the same for the sum of the gas and the
new formed stars. }
\label{conc0302}
\end{figure}

In the  simulation c2, the
stars produce a barred  feature,   lasting until  $z=0$ (Fig. \ref{dens2}).  
This is an important difference with the  case having the same gas fraction in
Paper 2, where the bar was decreasing its strength and disappearing at
$z \approx 0.15$.  
The same  behaviour arises also in the c3 case (Fig. \ref{dens3}), which corresponds to a larger initial
gas fraction: the bar is living until the final redshift, whereas, without the
star formation, the bar was disappearing at $z \approx 0.6$.\\
 The gas forms a bar since the beginning of the disk evolution and the new
stars originating from the gaseous component  
assume  an elongated structure which fuels the  old stars bar
feature:   the new stars  are dynamically cold and they strongly feel the
resonances in the equatorial plane, where they are born.
 In the simulations with
higher gas fractions, they are  swept out in the central regions
by vertical resonances which results in the formation of a central young 
bulge \citep{Friebe95}.  
\begin{figure}
\begin{center}
\includegraphics[width=7cm]{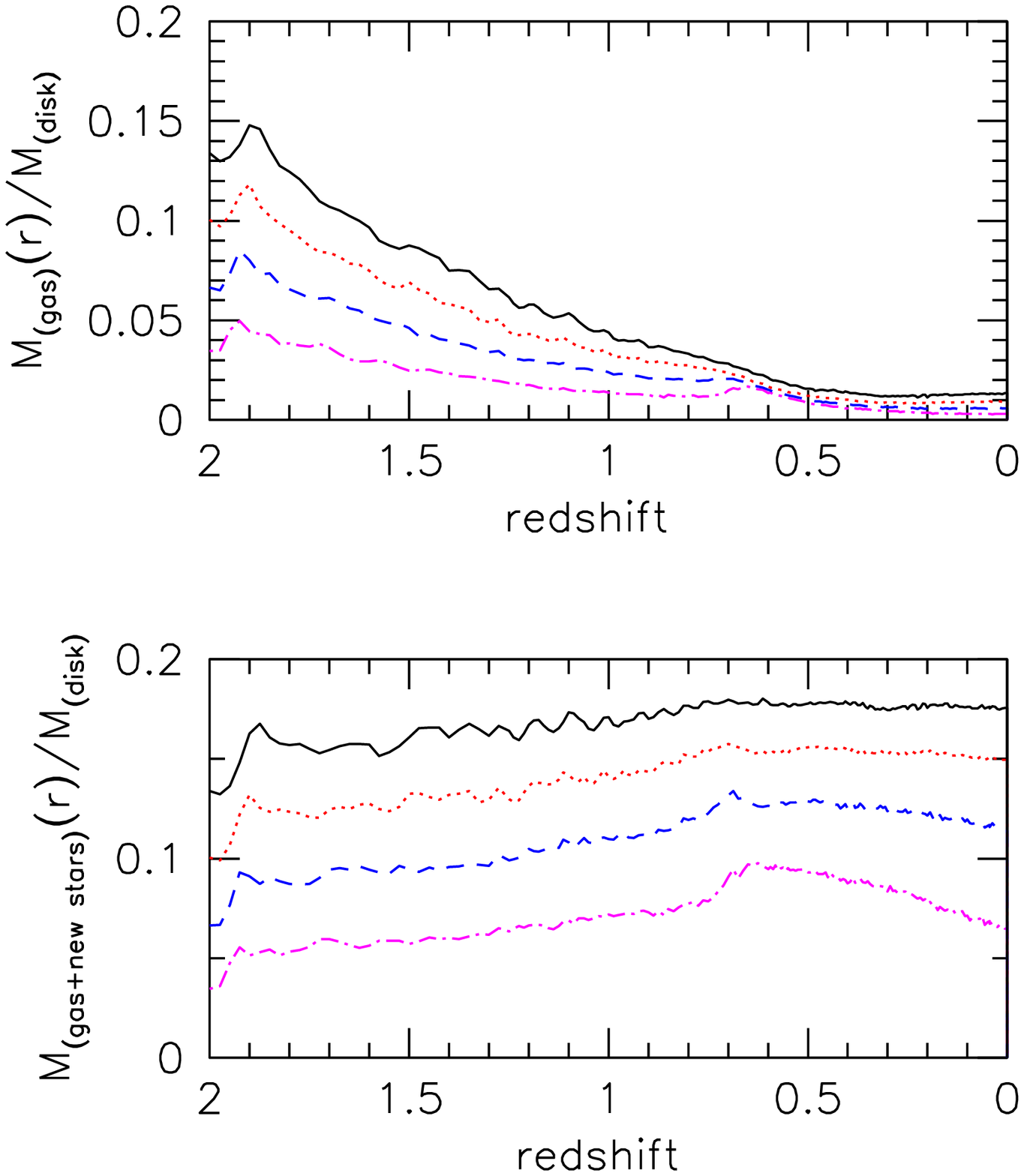}
\end{center}
\caption{Same as in Fig. \ref{conc0302} but for simulation c3.}
\label{conc0304}
\end{figure}
According to the claim of
\citet{ber98}, the growth of a gas mass
concentration in the disk dissolves the regular orbits in the stellar bar. 
In Paper 2, we  deduced a threshold value  for the gas concentration inside
a radius of 2 Kpc, 9\%, able to destroy the bar in the more massive disks. 
If the star formation is included, it  transforms the dissipative component into a non
dissipative one and therefore  the central concentration of the gas is lower.
Fig.  \ref{conc0302}  and Fig. \ref{conc0304} 
show the evolution 
of the gas mass in the inner regions, i.e., of the gas concentration (upper panel), 
and  that of the mass of gas plus new stars (lower panel) in the same regions, 
in simulations c2 and c3 respectively.
In both these cases, the gas mass inside 2 kpc never overcomes 
the threshold value pointed out in Paper 2. However,  the gas+new star 
concentration in simulation c3 overcomes this  threshold 
value at z=0.6, without any consequence on the bar
observed (Fig. \ref{strength_03}). 
\begin{figure}
\begin{center}
\includegraphics[width=7cm]{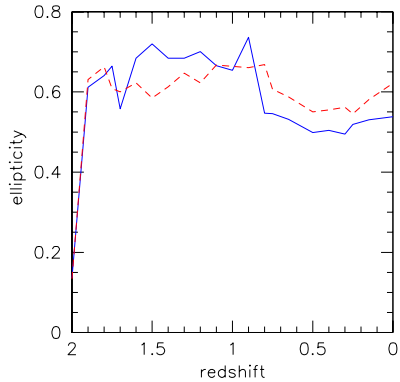}
\end{center}
\caption{Evolution in redshift of the ellipticity  for 
 simulations c2 (blue full line)  and c3 (red dashed line).
}
\label{strength_03}
\end{figure}
\begin{figure}
\begin{center}
\includegraphics[width= 6cm]{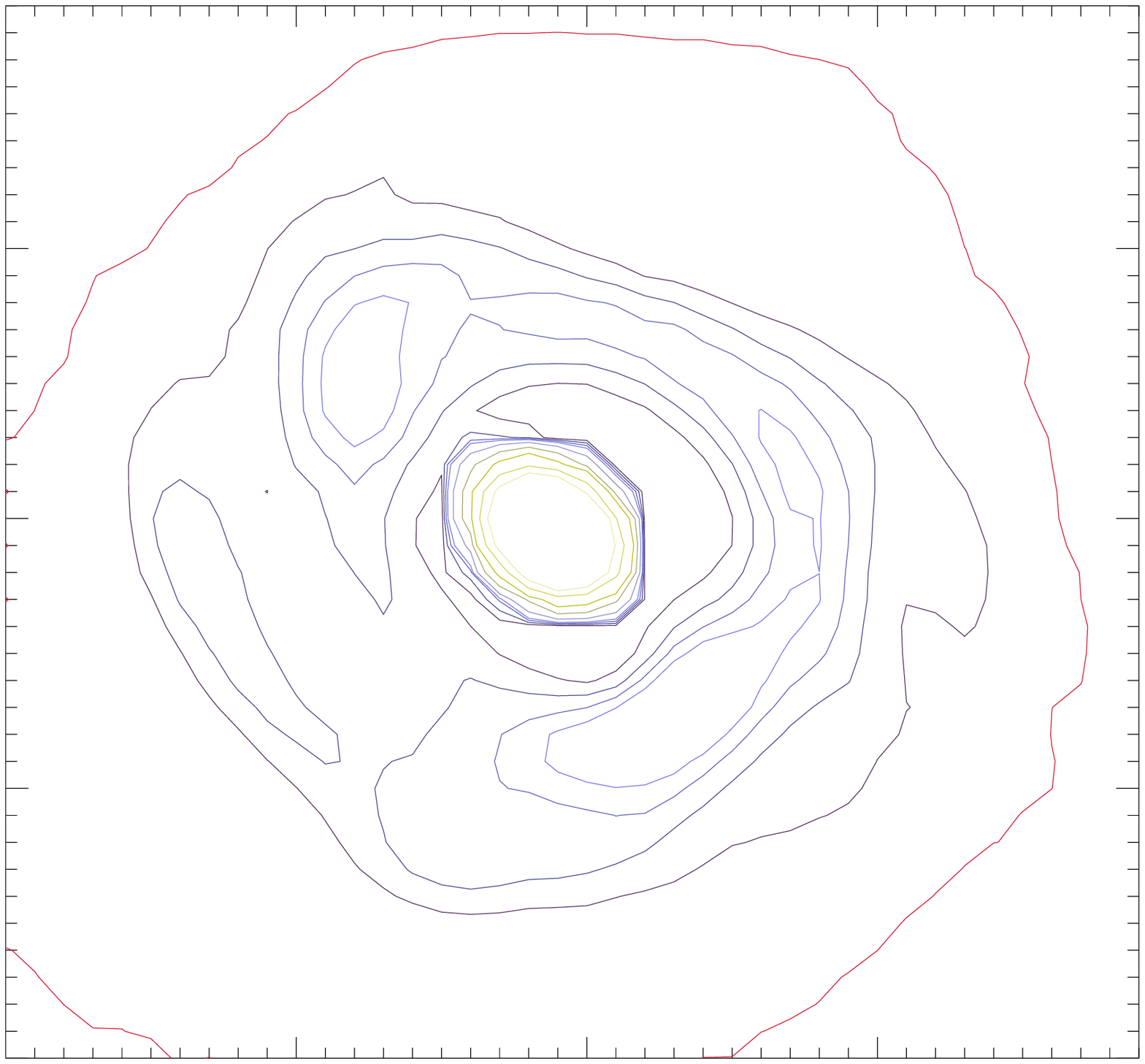}
\includegraphics[width= 6cm]{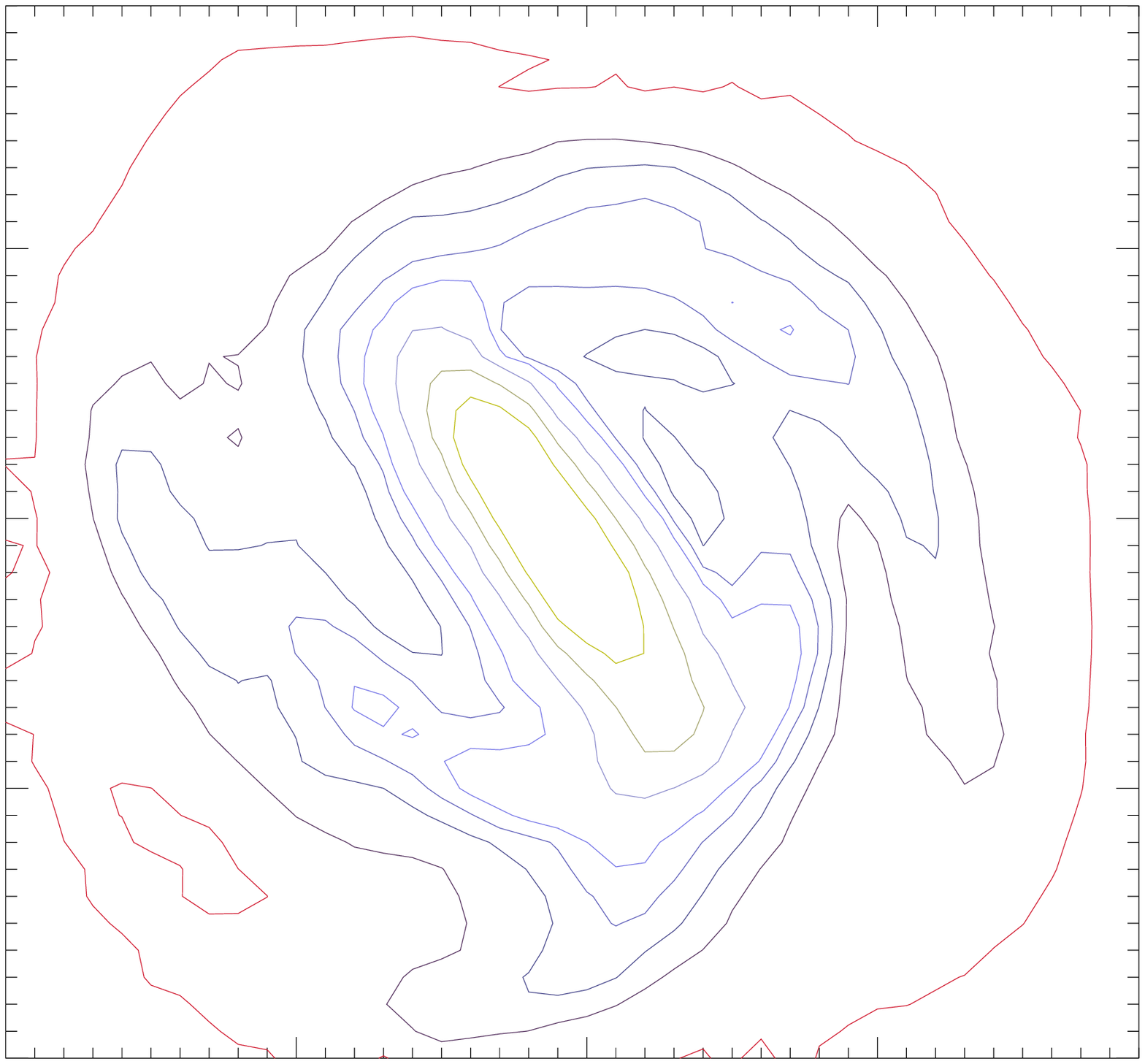}
\end{center}
\caption{Gas morphologies  at z=1 for simulations c1 (left panel) and c3 
(right panel).}
\label{gas_z1_03}
\end{figure}
\begin{figure}
\begin{center}
\includegraphics[width=7cm]{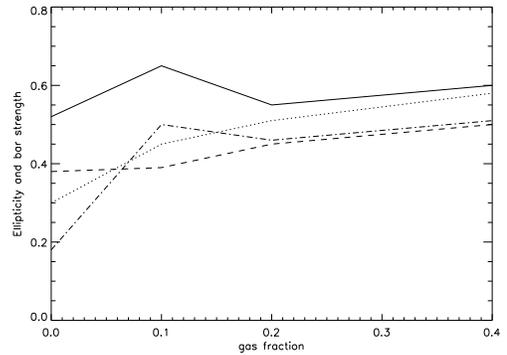}
\end{center}
\caption{ Behaviour of the bar strength and ellipticity at z$=0$ for our set of
cosmological simulations  with increasing gas fraction. 
 $Q_b$ (dotted line) and ellipticity (full line) of our
more massive disks (i.e. disk--to--halo mass ratio
0.33); $Q_b$ (dot--dashed line) and ellipticity (dashed line) of our
less massive, DM-dominated, disks (i.e. disk--to--halo mass ratio
0.1).
}
\label{strength}
\end{figure}
This is because the effect of bar dissolution due to the central mass 
concentration  cannot be achieved including the mass of new stars which
are crossing  the central region through elongated orbits and are not really 
concentrated as the gas.
In the simulation c3 the gas shows more developed  arms at z=1 than in
the case c1 (Fig. \ref{gas_z1_03}). 
\begin{figure}
\begin{center}
\includegraphics[width= 7cm]{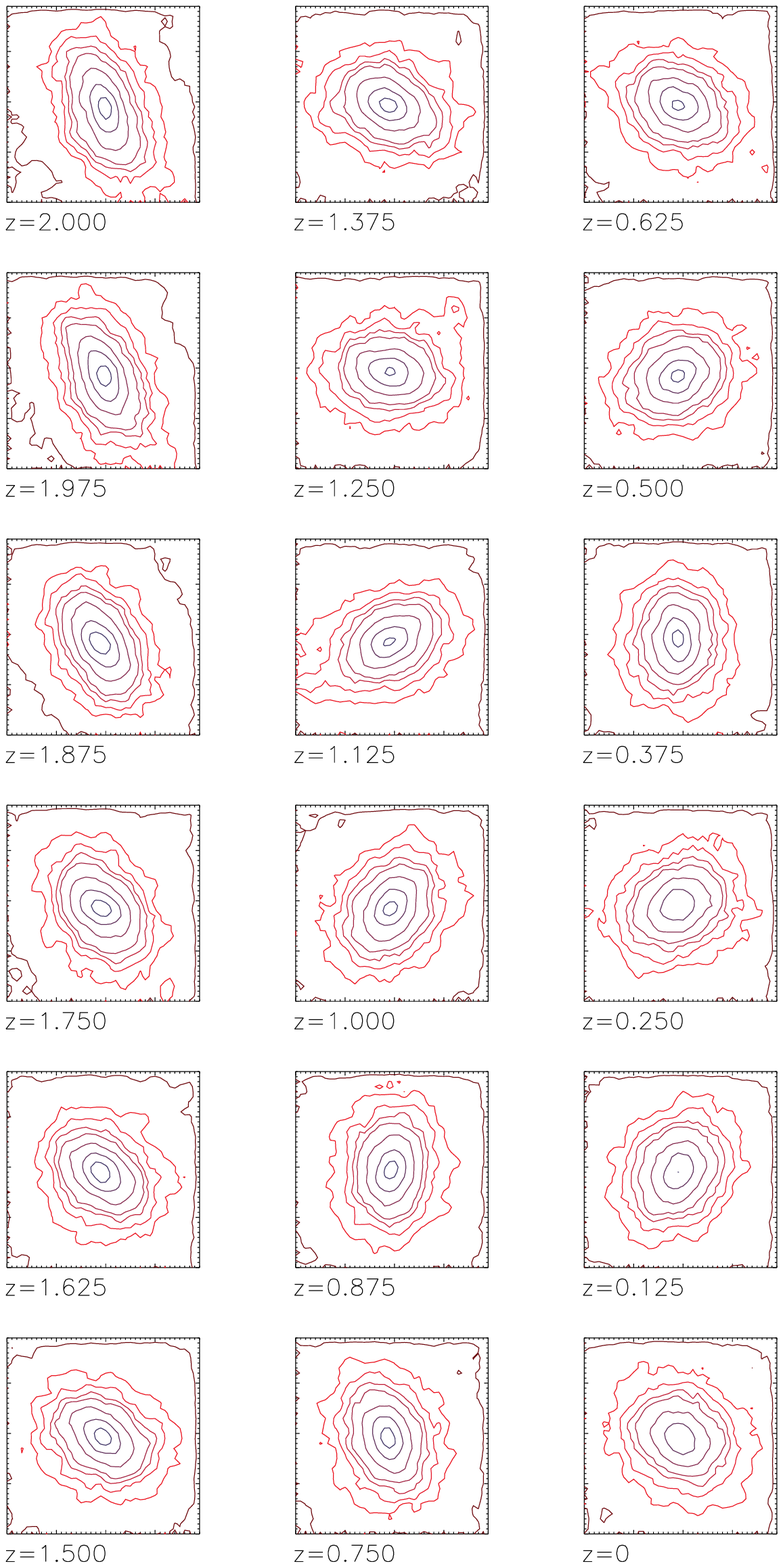}
\end{center}
\caption{ Isodensity contours of the DM in the xy plane, from redshift $z=2$
  to 
$z=0 $ for the simulation c4. Density levels and spatial resolutions 
are the same as in Fig. \ref{evol1}. The syncronicity of the semimajor axis of
the halo with the one of the inner old stellar bar can be noticed by direct comparison of each
panel (after z=1.8) with the corresponding one in Fig. \ref{evol2}. }
\label{dm_bar1}
\end{figure}
\begin{figure}
\begin{center}
\includegraphics[width= 7cm]{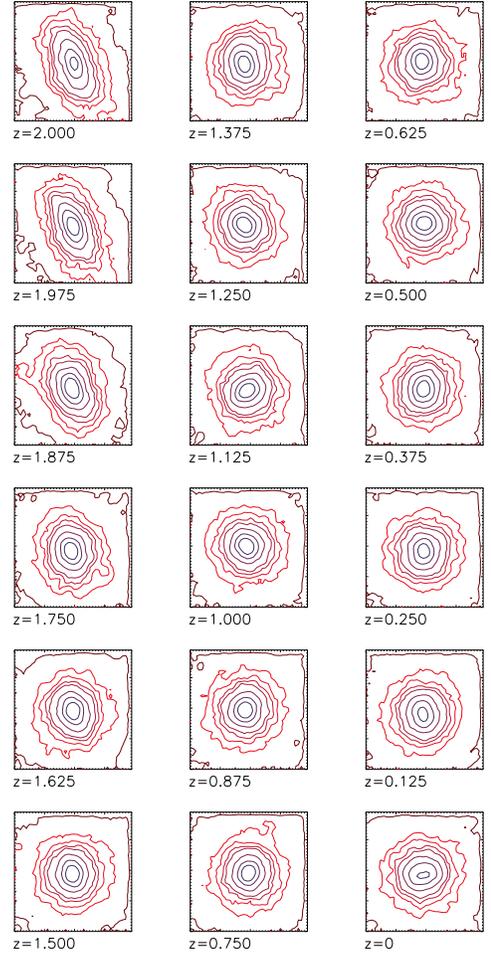}
\end{center}
\caption{ Isodensity contours of the DM, as in the previous figure, for the  simulation c1.}
\label{dm_bar2}
\end{figure}
\begin{figure}
\begin{center}
\includegraphics[width=6cm]{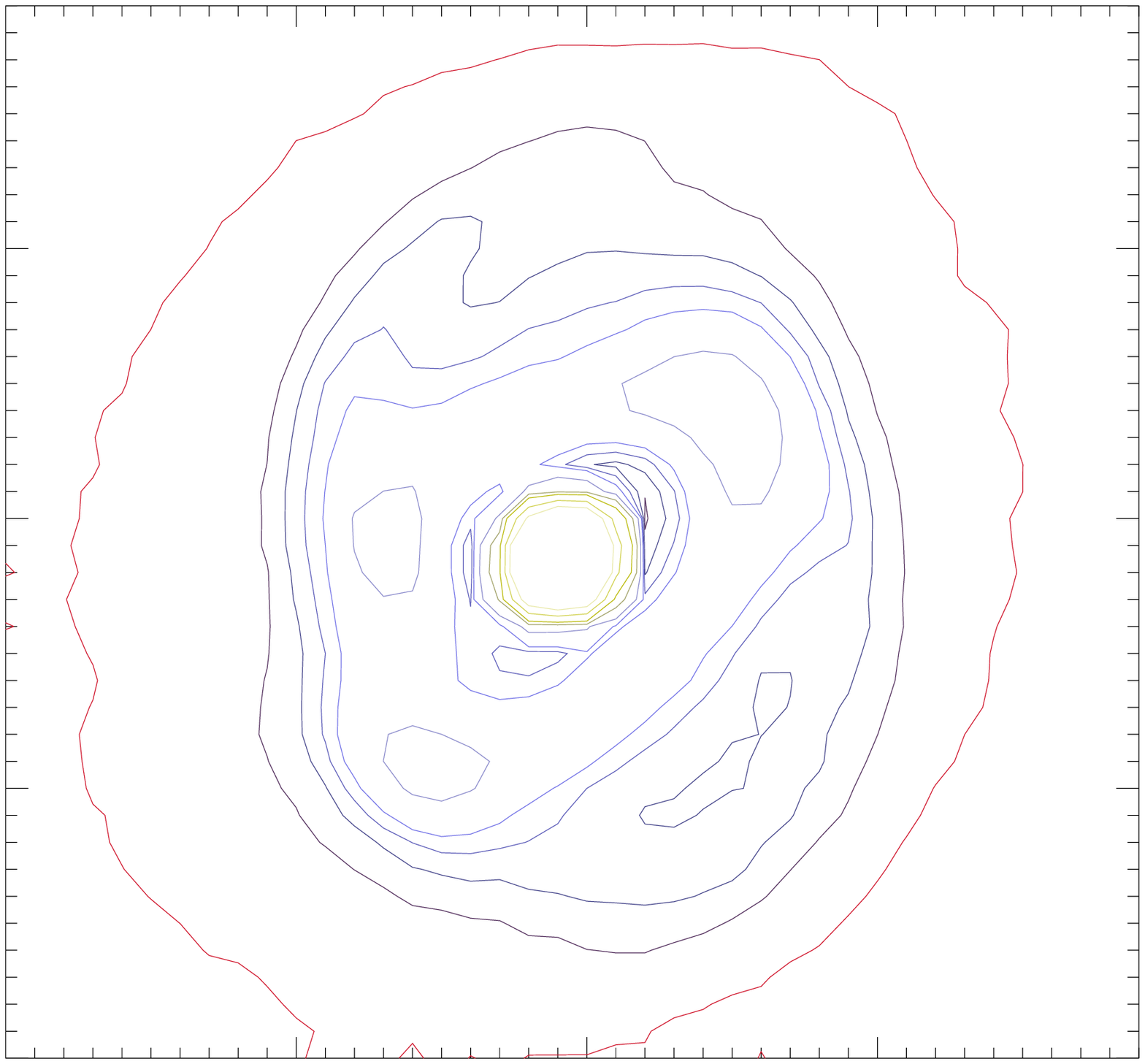}
\includegraphics[width=6cm]{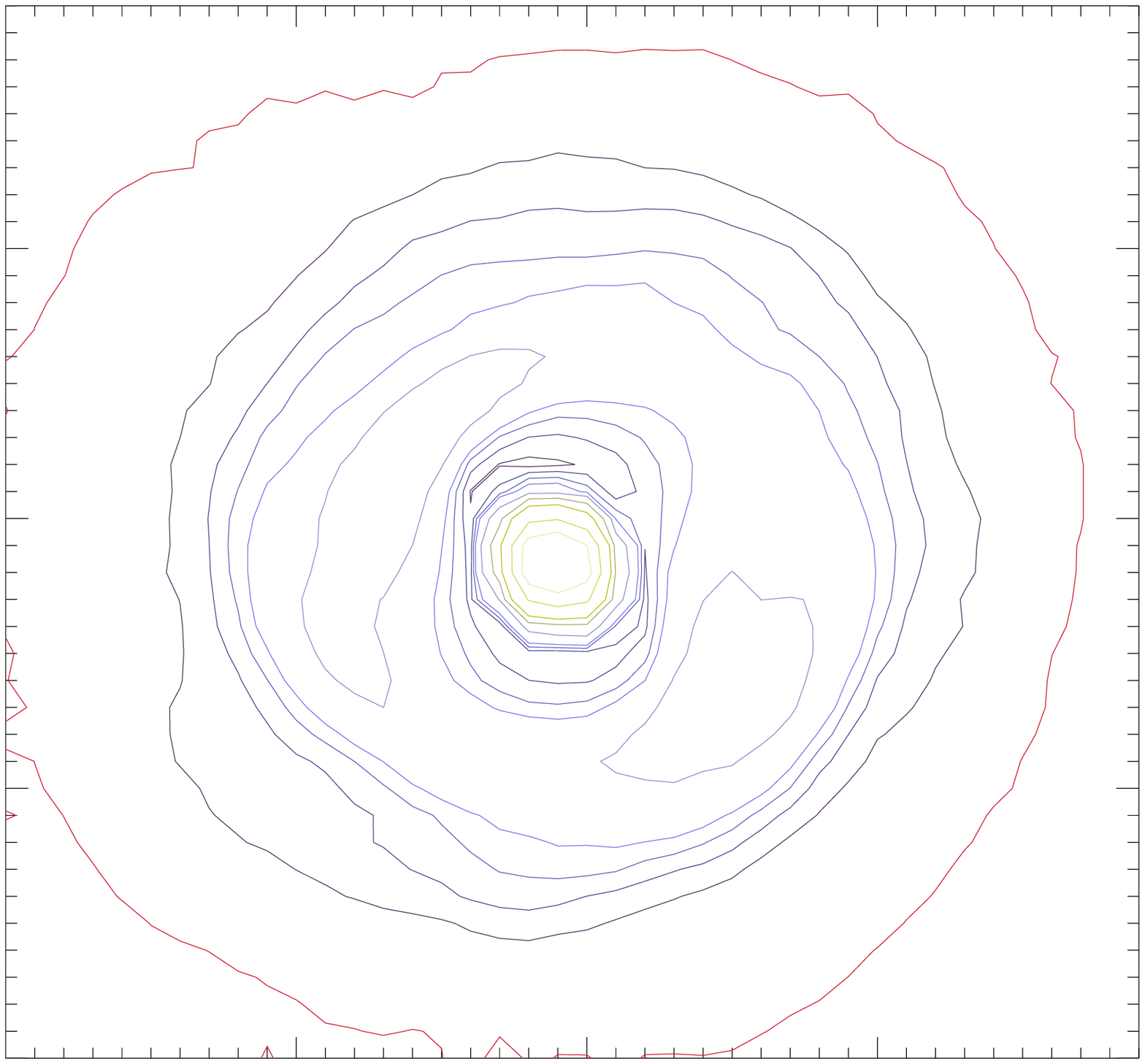}
\end{center}
\caption{  Same as in Fig. \ref{gas_z1_03} but for simulations c4 and c6.}
\label{gas_z1_01}
\end{figure}
But this feature, which is
considered as a possible mechanism for the bar destruction has no impact on
the bar in these new simulations \citep{Bou05}. 

A different discussion is devoted to the DM-dominated disks, which, as we already
stressed in Paper 1 and 2, show bars which are not a classical product of the
self-gravity or of angular momentum exchanges, but they are features that
strongly depend on the dynamical state and evolution of the cosmological halo,
 and we did not observe their
distruction even by an higher gas concentration.
\begin{figure}
\centering
\includegraphics[width=7cm]{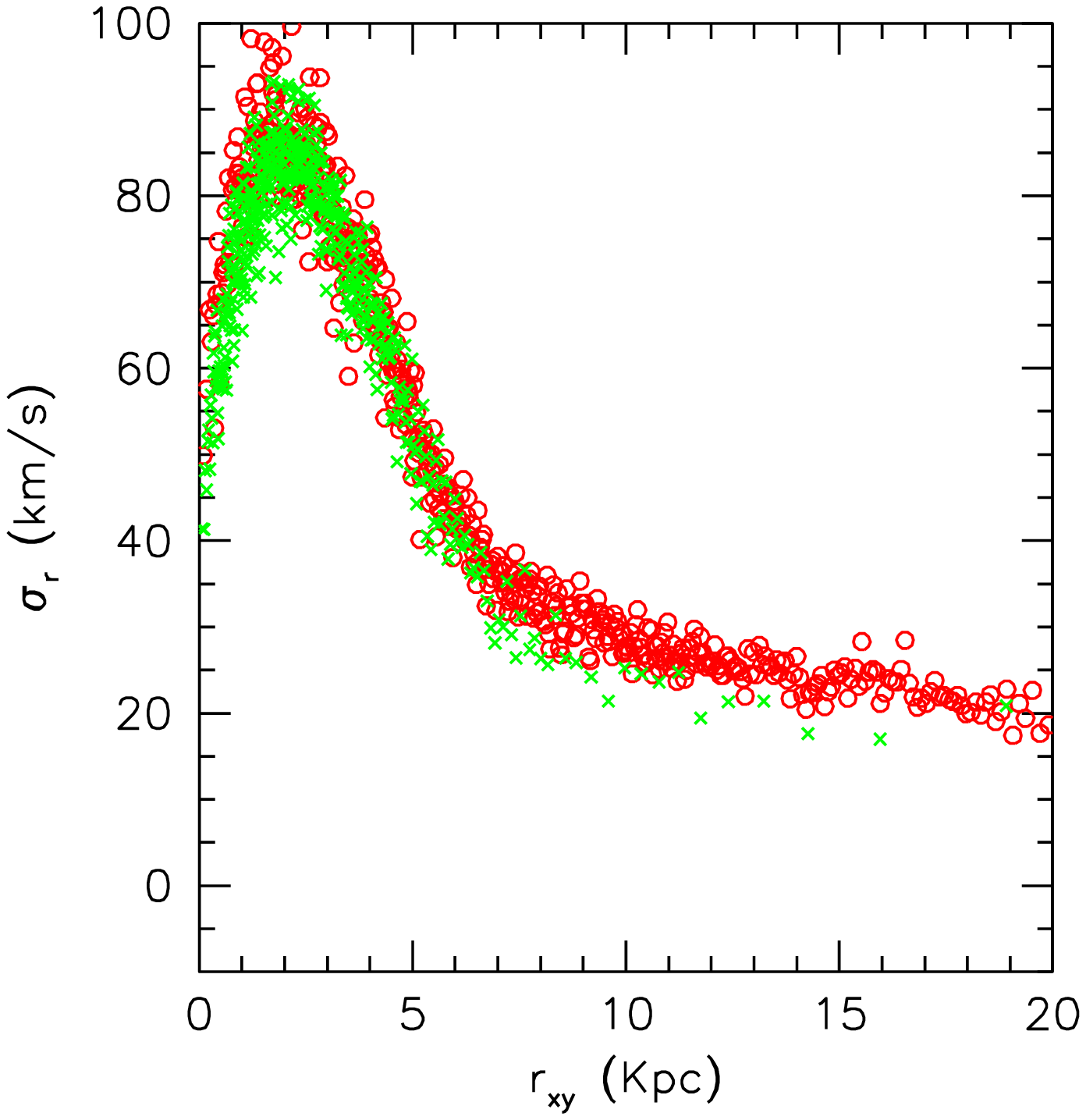}
\includegraphics[width=7cm]{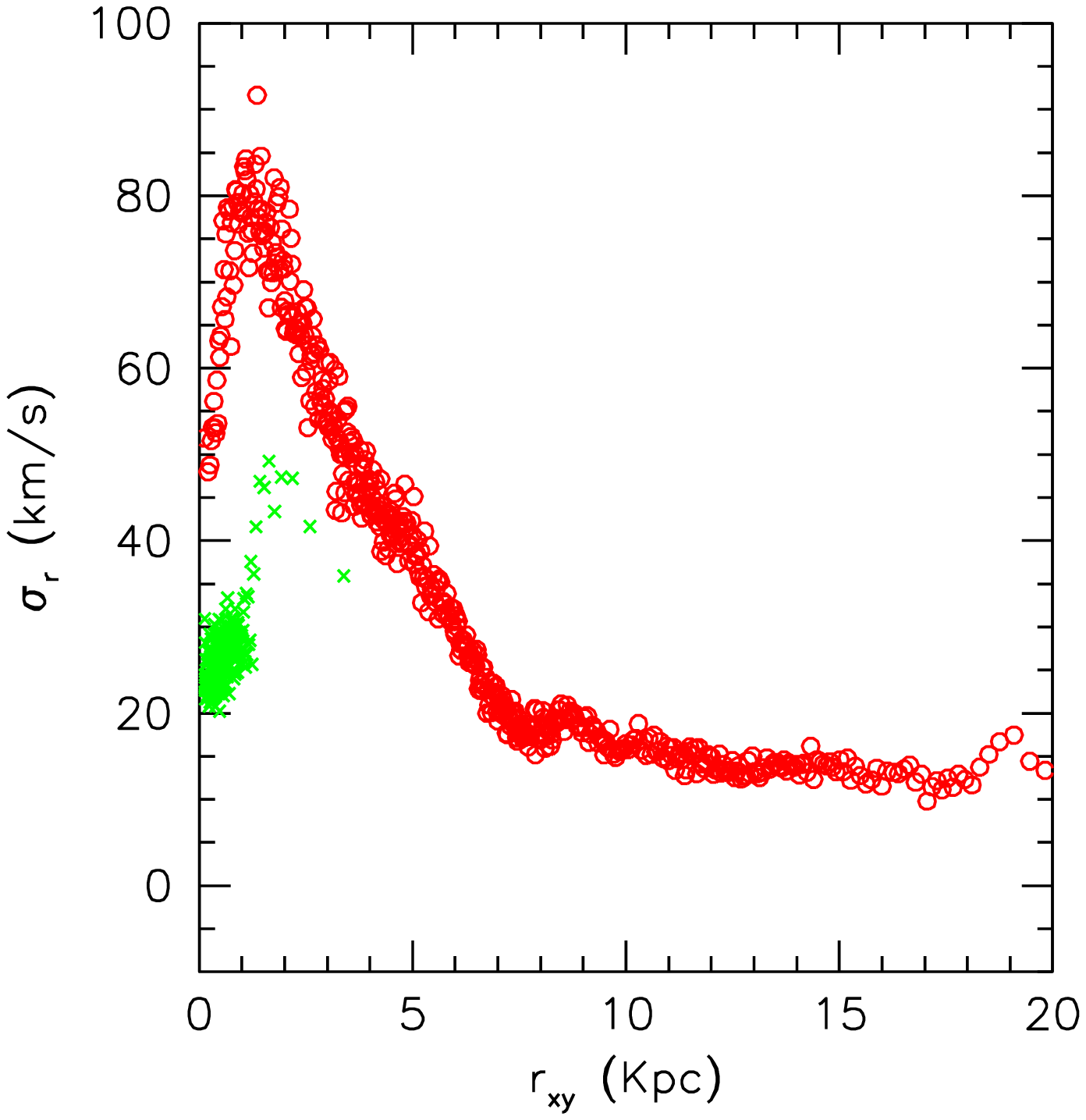}
\caption{Radial velocity dispersion at z=1 of old stars (red open circles) and new
  stars (green crosses) for simulations c2 (left panel) and c5 (right panel).  }
\label{dispersion}
\end{figure}
Old stars in simulation c4  show a stellar bar stronger than that
 in the pure  stellar disk  of the same
mass (ellipticity 0.39  instead of 0.3,  see  Tab. \ref{fin_table} ) but weaker than
that arising from the simulation with the same disk mass and gas fraction
but without star formation  (Paper 2).
Bar-in-bar features are present along the whole evolution (simulation c4) 
until z=0  (Fig. \ref{dens4}). 
Simulations  c5 and c6, which correspond to increasing gas 
fractions, show 
 similar features during the system evolution. In this set  of DM dominated
 disk simulations, the  final bar is increasing its
 strength with the increasing gas fraction (Fig. \ref{strength}, Tab. \ref{cosmsimtable}). 
  This trend is different from the one we observed in  Paper 2, where  
 the increasing gas fraction produces an increase to a maximum and then  a 
 decrease of the bar strength.\\
 In DM dominated disks the  bars in the old star component are the 
 imprint of the halo  cosmological properties, as discussed in the previous
 Papers.  Fig. \ref{dm_bar1}, shows the isodensity contours
  of the halo in the plane of disk in the same frame, with the same 
  isodensity levels, and box sizes as those of the old stars in Fig. \ref{evol1} and Fig.\ref{evol2}.    
  The semimajor axis of the triaxial DM halo, for $z < 1.8$ is both in phase with the
  central bar of the disk  and they have the same pattern speed.
   
  In the more massive disks (Fig. \ref{dm_bar2}), the central DM region 
  becomes more round, due to impact of the disk itself, and the coupling with
  the stellar bar disappears. Therefore, in the case of the lighter
    disks,  the cosmological halo properties  drive the formation of the bar, 
    whereas in the more massive disks, the bar is a classical product of
    the disk instability.
In the DM dominated disks the new  stellar component produces
a small spheroidal
bulge which has a disky shape in the  simulation c6.  The gas component is  
confined in the central regions from the beginning of the evolution, and 
therefore the new stellar
component arranges itself in a
small spheroidal bulge. 
Therefore, in DM dominated disk, the new stars at z=0 have no
bar features, at difference with  more massive disks.
  We recall that,  in DM dominated disks the bars are not due to
classical resonances.  So, the new stars do not feel these
resonances whereas this is what happens in the cases of the more massive disks.
  The decoupling between
  new and old stars is depicted also by 
  In Fig. \ref{dispersion}  the 
  distribution of radial velocity dispersions in a massive
  disk at z=1 is compared with the one in a DM dominated disk at the same 
  redshift;  in  such a case the decoupling between
  new and old stars is well depicted.
  The radial velocity dispersion of this new star
  component is lower than that of the old stars. 
  At the end of the evolution we  therefore see a small bulge of new stars
hiding a small nuclear bar of old stars.  Such a
nuclear bar remains, as observed in Papers 1 and 2
and  stressed by Fig. \ref{dm_bar1}, as a genuine product of the cosmology, 
since the light disks are classically below the instability threshold.

\subsection{Bar pattern speed}
\begin{figure}
\begin{center}
\includegraphics[width=7cm]{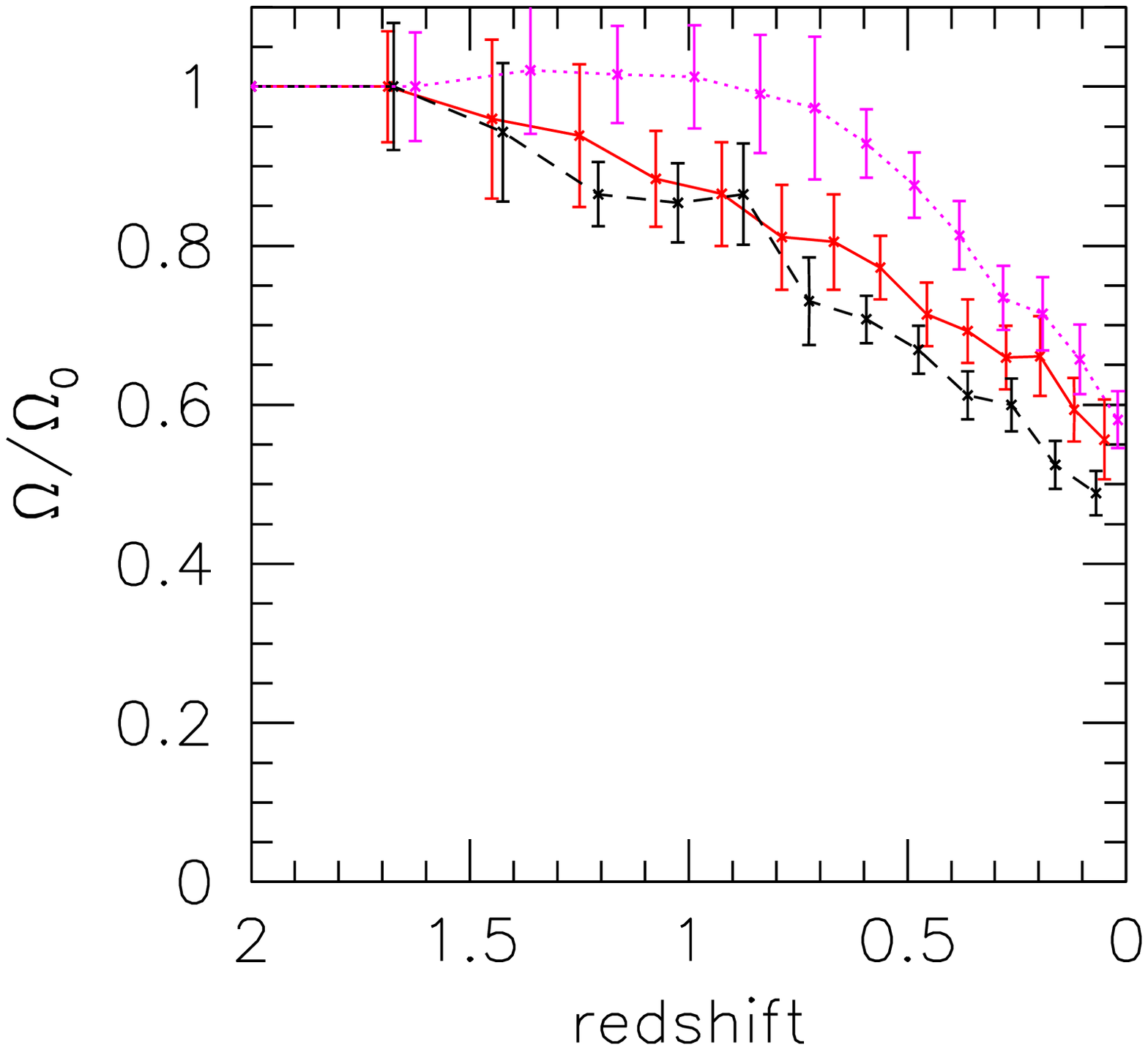}
\end{center}
\caption{ Evolution of the bar pattern speed with the redshift for simulation
  c1 (black dashed line, c2 (red full line) and c3 (magenta dotted line).}
\label{pattern1}
\end{figure}

\begin{figure}
\centering
\includegraphics[width=7cm]{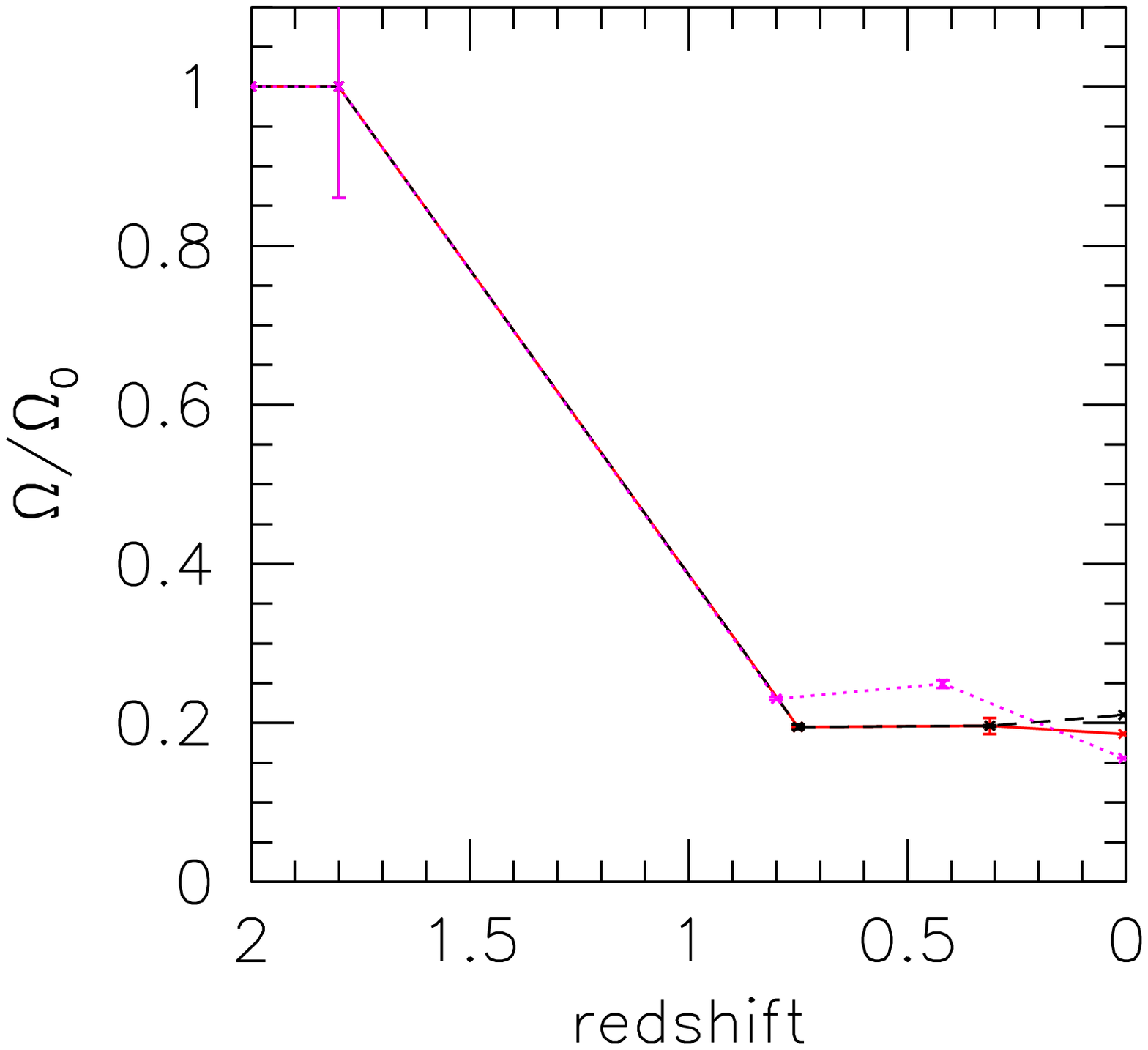}
\caption{ Evolution of the bar pattern speed with the redshift for simulation
  c4 (black dashed line), c5 (red full  line) and c6 (magenta dotted line).}
\label{pattern2}
\end{figure}

The bar pattern speed, $ \Omega_p$ is the angular speed  of the
 bar-like  density wave as viewed from an inertial 
frame.  Here, it is evaluated following the position angle of the major bar
axis during the  disk evolution,  where the major axis is selected with the
same criterium E1 quoted in \citet{MiWo}, namely as the radius where the
ellipticity profile reaches a maximum as in Paper 2.   The values of  $ \Omega_p$ as function of the redshift
  for  the more massive disks and
  the lighter disks are shown in 
  Fig. \ref{pattern1}  and Fig. \ref{pattern2} respectively.

For the more massive disks,  we observe a  decrease of   $ \Omega_p$ in 
all the  simulations. This is a different behaviour from the one observed
 in Paper 2: there we noticed  an increase of the pattern speed coupled
 with the increasing gas fraction, since the
gas presence both shortens the bar and decreases its ellipticity.   

\begin{figure}
\centering
\includegraphics[width=7cm]{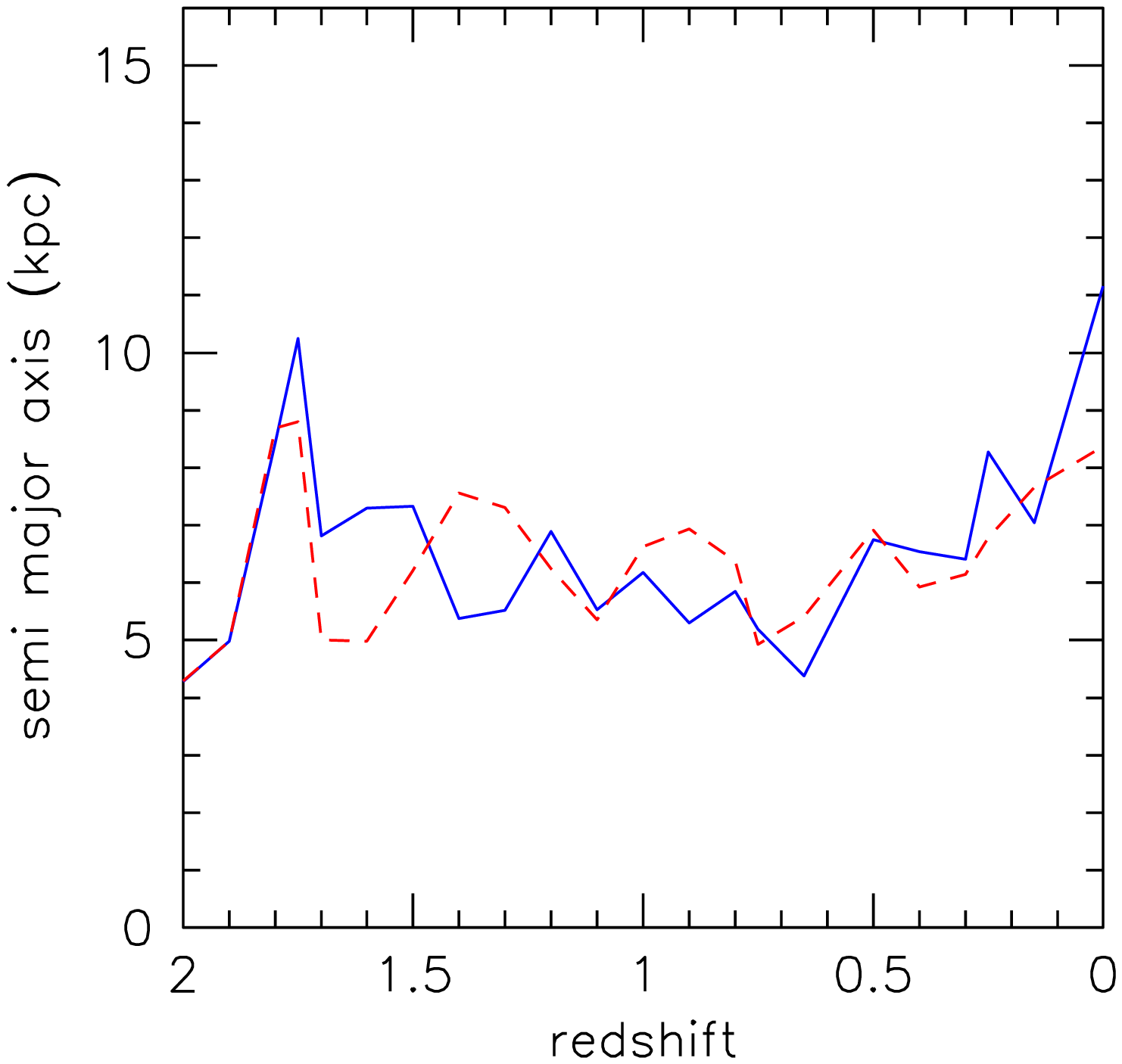}
\caption{ Behaviour of the semimajor axis for the simulations c2
  (blue full line) and c3 (red dashed line).}
\label{maxmin}
\end{figure}

This mechanism is no longer working when the star formation is activated, since
the ongoing star formation reduces the gas concentration
in the central regions and the new stars arrange into an elongated structure
which enforces also the length of the original bar (Fig. \ref{maxmin}, see
also the previous section).

For the DM dominated disks the evaluation of $\Omega_p$  in this paper
as in the previous ones, is done by referring to
  the angular velocity of the long-lived  central bar.
Its  behaviour is the same  as observed in Paper 2,  namely it
 shows a fast decrease before
 redshift $z\simeq 1$, and then a quasi stationary behaviour until redshift 0.

\section{Discussion and conclusions}

We presented six  cosmological simulations  with the same  disk--to--halo mass
ratios   as  in  Paper 1 and Paper 2. In order to study the impact of the 
forming stars from the gaseous   component, here we included and 
varied its percentages inside disks of different disk-to-halo mass ratios, 
as in  Paper 2 where the star formation was switched off.  In Table
\ref{fin_table} we show the final data for
ellipticities and semimajor axes resuming the
results of our three papers.

The old star component show  a long lasting bar, 10 Gyr old, in
all the simulations  of this work.
\begin{table*}
\caption{Cosmological disks simulations: global results}
\label{fin_table}
\centering
\begin{tabular}{c c c c c c }
\hline\hline
Disk mass& gas fraction & {$   $} no star formation & {}& star formation & {}\\
{}& {}  & ellipticity& $a_{max}$ & ellipticity & $a_{max}$ \\
\hline
0.33 & 0. &  0.52 & 5 & - & - \\
0.33&  0.1 &  0.68 & 8.4 & 0.65 & 8 \\
0.33& 0.2 &  0.1 &  no bar  & 0.55 & 11 \\
0.33& 0.4 & 0.07 &  no bar & 0.6 & 8.4 \\
0.1& 0. &  0.3 & 6.5 & - & - \\
0.1&0.1 &  0.58 & 5.8 & 0.39 & 3 \\
0.1& 0.2 &  0.6 & 5.4  & 0.45 & 3 \\
0.1&  0.6 & 0.42 & 5.8 & 0.5 & 3 \\
\hline
\end{tabular}
\end{table*}
Moreover, in all such simulations,  except for simulation c6,  
the  bar is stronger than that developed in the pure stellar case
with the same disk-to-halo mass ratio 
but weaker than that formed in the case of the same gas fraction without star formation.
 
\begin{figure}
\begin{center}
\includegraphics[width= 7cm]{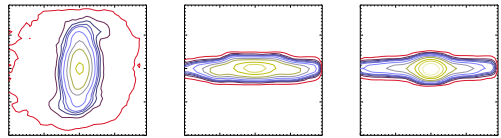}
\end{center}
\caption{ Isodensity contours of the old stars at z=0, for the  simulation c2,
rerunned with the Katz'96  star formation recipe}
\label{katz}
\end{figure}

We find that the star formation, reducing the central gaseous mass
concentration, allows the bar to survive until the end of the evolution 
also in the more massive disks, at
variance with the results in Paper 2 (where the gas was not allowed to
form stars):  in such a case a gas fraction  0.2  was able to destroy the 
bar.
Even if  some
details of the disk morphologies  obviously depend on the star
formation prescription, the  bars arising here  in the c2 and c3
simulations are   due to the
reduced central gas concentration and thus to the 
weakening of its effect on the bar itself. 
We verified this point  by rerunning simulation c2 changing the star
formation prescription.  Instead of the GADGET-2 effective model, we used a
simple algorithm in which, when a gas particle reaches a given density
threshold  with a temperature lower than a minimum temperature threshold, 
it is converted into a star particle, as in  \citet{Katz96}. Fig. \ref{katz}
shows the morphology of the old stellar component 
for such a  different star formation recipe:
 the survival of the bar is due to the star formation, independently on
its details.\\
In all the simulations of more massive disks, 
the new stellar component
shows a barred shape coupled with that of old stars but
with semimajor axis and ellipticity values smaller than those.
In  all the simulations  of DM dominated disks, a  bar feature is
maintained in  the old stellar component at the disk center,
and the new stars form a spheroidal
bulge that  hides the  bar of the old stars.  This effect could be of
interest as far as the  observations about the Milky Way are concerned
where, in addition to the main bar, a small nuclear
bar consisting of old stars seems to be included in the bulge 
(see e. g. \citet{ala01}).

The final  strenght of these old star bars  increases by  increasing 
gas fraction  and their pattern speed  is quickly decreasing before z$=1$.

The classical results obtained outside the cosmological scenario are
no longer applicable. 
This conclusion  remarks the results of Paper 1, where it was shown that
in the DM dominated disks the bar feature is triggered and maintained by the 
cosmological  properties of the halo (namely its triaxiality and its 
dynamical state).

From the models presented here we suggest that a very low pattern speed (few $
Km\, s^{-1}\, Kpc^{-1}$) could be a  signature 
of  a dominating halo and not a classical product of the disk instability. 

The whole set of cosmological simulations we  presented here is 
suggesting that the star formation works in favour of maintaining the bar
feature in self gravitating disks,  against the effect of increasing gas
fractions,  whereas in  DM  dominated disks,
it slightly reduces the bar strength. Moreover, inside such  disks,
the new  stars  are arranged in non barred systems which are decoupled 
from the  bar  of the old star component.

{\bf Acknowledgements}  
Simulations have been performed on the CINECA IBM CLX Cluster, thanks
to the INAF-CINECA grants cnato43a/inato003 ``Evolution of disks in
cosmological contests: effect of the star formation inside the disk'', and on the Linux PC Cluster of
the Osservatorio Astronomico di Torino. We wish to thank   V. Springel for
kindly providing us with his code GADGET, and Martina Giovalli which
implemented the simple star formation prescription in it.  We thank the referee for his
suggestions, useful to improve the paper.

\bibliographystyle{aa}
\bibliography{5286data}
\end{document}